\documentclass[11pt]{iopart}

\usepackage[dvips]{color}

\newcommand{\bm}[1]{\mbox{\boldmath $#1$}}

\def\Journal#1#2#3#4#5#6{#5 #6 {#1} {\bf #2} #3#4}

 \def\JGP{\em J. Geom. Phys.}

\def\GRG{\em Gen. Rel. Grav.}
 \def\CMP{\em Commun. Math. Phys.}
\def\PRL{\em Phys. Rev. Lett.}
 \def\AIHP{\em Ann. Inst. H. Poincar\'e}
\def\PPLLA{{\em Phys. Lett.} A }

\newtheorem{prop}{Proposition} 
\newtheorem{theorem}{Theorem} 
\newtheorem{corollary}{Corollary}[theorem]

\def\fin{\hfill \rule{2.5mm}{2.5mm}\\ \vspace{0mm}}
\def\finn{\hfill \rule{2.5mm}{2.5mm}}

\def\spacetime{(\mathcal{M},g_{\mu\nu})}
\def\orbits{(\Sigma_3,\hh_{ab})}
\def\hh{\widehat h} 
\def\RR{\widehat R} 
\def\hhd{\hat \nabla} 
\def\hheta{\hat \eta} 

\def\ernst{\mathcal{E}}
\def\eee{X}
\def\ff{\mathcal{F}}

\def\sss{\Sigma}

\def\yy{\mathcal{Y}}
\def\lll{L}
\def\genj{\iota}
\def\opp{\widetilde}
\def\oppRe{\Re}
\def\oppIm{\Im}
\def\cxc{\overline}
\def\cxRe{\mbox{Re}}
\def\cxIm{\mbox{Im}}
\def\jc{\breve}
\def\teq{\acute}

\def\nn{N}

\def\trian{\delta}
\def\aa{n}

\def\setrules{Rules}

\def\fa{\left(\alpha_1-\frac{2}{r}\alpha\right)}
\def\fc{\left(\gamma_2-\frac{2}{r}\gamma\right)}

\begin{document}

\title{A local characterisation for static charged black holes}

\author{Guillermo A Gonz\'{a}lez$^{1}$ and Ra\"{u}l Vera$^{2}$}
\address{$^{1}$Escuela de F\'isica, Universidad Industrial de Santander, A. A.
678, Bucaramanga, Colombia}
\address{$^{2}$Dept. F\'isica Te\'orica, Universidad del Pa\'is
Vasco/Euskal Herriko Unibertsitatea, 644 PK, 48080 Bilbao, Basque Country, Spain}
\ead{guillego@uis.edu.co, raul.vera@ehu.es}

\begin{abstract}
We obtain a purely local characterisation that
singles out the Majumdar-Papapetrou class, the near-horizon
Bertotti-Robinson geometry and the Reissner-Nordstr\"om exterior solution,
together with its plane and hyperbolic counterparts,
among the static electrovacuum spacetimes.  
These five classes are found to form
the whole set of static Einstein-Maxwell fields
without sources and conformally flat space of orbits,
this is, the conformastat electrovacuum spacetimes.
The main part of the proof consists in showing
that
a functional relationship between the
gravitational and electromagnetic potentials must always exist.
The classification procedure provides also an improved characterisation
of Majumdar-Papapetrou, by only requiring a conformally flat space
of orbits with a vanishing Ricci scalar of the usual conveniently
rescaled 3-metric.
A simple global consideration allows us to state that
the asymptotically flat subset of the Majumdar-Papapetrou class
and the Reissner-Nordstr\"om exterior solution
are the only asymptotically flat conformastat electrovacuum spacetimes.

\end{abstract}



\section{Introduction}
The (standard) Majumdar-Papapetrou and
Reissner-Nordstr\"om metrics
are known to describe, under rather general conditions,
the exterior geometries of the
static charged black holes, as shown by the recent uniqueness
theorems (see \cite{chrus_tod_mBH} and references therein).
The aim of this work is to provide an essentially local characterisation
for the Majumdar-Papapetrou class and the Reissner-Nordstr\"om
exterior solutions.
Local characterisations are important, not only for being essential ingredients
for the improvement of the global charaterisations of black holes provided
by the uniqueness theorems,
but also for a better understanding of the solutions
and its potential use in stability problems.
We first find a purely local uniqueness result
that characterises Majumdar-Papapetrou, the near-horizon geometry
and Reissner-Nordstr\"om, together with its plane and hyperbolic
counterparts, among the (strict) static electrovacuum spacetimes.
Global considerations can then be used to
restrict the set conveniently.

The known characterisations of the different families of black hole
metrics vary from the (purely) local, the ``essentially'' local and
those of global nature in quite a gradual manner.
Whether or not some global property is preferable to some stronger
local constraints is not clear `a priori'
(see e.g. the discussion about some
possible Kerr characterisations included in \cite{MARCKerr2}).
Nevertheless, it is always convenient to try to minimise the number
of constraints present in any given local characterisation. 
Take the different characterisations
we have for the Schwarzschild metric. Although Birkhoff's theorem
constitutes a nice and purely local characterisation, it seems of no use in the
uniqueness theorems.
Another purely local characterization
which involves only the Weyl tensor and the metric itself
is given in \cite{ferrSch}. 
A more convenient characterisation ingredient appears to be
the conformal flatness of the hypersurfaces of constant static
time (conformastat), since that constitutes a crucial step in the uniqueness
theorems as they stand now.
This is, in fact, not purely local, since this
characterises Schwarzschild among the static and \emph{asymptotically flat}
vacuum spacetimes.
Indeed, conformastat vacuum spacetimes 
comprise three \cite{Das71} (see also \cite{perjes0})
out of the seven families that constitute the whole set of degenerate (type $D$)
static vacuum spacetimes \cite{EhleKundt,sols}.
These three families correspond to the Schwarzschild solution
together with its plane and hyperbolic counterparts
(Class $A$ in Table 18.2 in \cite{sols}).
The Schwarzschild solution
can be singled out by requiring asymptotic flatness.
This is a simple global consideration, and in this respect
one may think of this as being an essentially local characterisation.

A natural step to follow is 
the generalisation of the above to static charged black holes.
Indeed, global arguments in the uniqueness theorems establish, again,
conformal flatness of the hypersurfaces of
constant static time \cite{heusler}. It is important to note, however,
that the same global arguments also imply that the gradients of the
gravitational and the electromagnetic potentials are aligned,
or in other words, that
the
potentials are functionally related.
It is these two facts, together with a ``non-degeneracy'' restriction
and asymptotic flatness, that lead eventually to spherical
symmetry and thus to the standard
uniqueness results for the non-extreme Reissner-Nordstr\"om black hole
(see e.g. \cite{heusler}).
One question we address in this paper is up to which extent
the alignment and the ``non-degeneracy'' properties can be relaxed
in a local characterisation. We believe this may be of use
on the improvement of the recent uniqueness theorems of (multi) black holes
(see \cite{chrus_tod_mBH} and references), since
we provide an essentially local
uniqueness result for the static charged black hole solutions,
binding together the Majumdar-Papapetrou
to the exterior Reissner-Nordstr\"om solution.

The characterisation we present here come by
finding the complete solution of the Einstein-Maxwell field equations
without sources for static spacetimes with a conformally flat space of orbits.
The only extra assumption
made is that the electromagnetic field inherits the symmetry, so that
it is also stationary.
We call such solutions conformastat electrovacuum spacetimes.
Note that we take the definition in \cite{sols} as standard,
following the original terminology by Synge \cite{synge}:
conformastationary are those stationary spacetimes
with a conformally flat space of orbits
and the conformastat comprise the static subset.

Conformal flatness corresponds to the vanishing
of the Cotton tensor associated to the
induced metric on the space of orbits. 
A general study of conformastationary spacetimes would follow
then an analogous path to the chatacterisation of the Kerr
and Kerr-Newman families of black holes among the stationary solutions.
In the Kerr case the crucial local property is the vanishing of the complex
Simon tensor \cite{walterKerr},
which generalises the Cotton tensor on the space of orbits.
The characterisation of
the Kerr metric in \cite{walterKerr} 
comes as a result of the equivalence
of the multipole structure of Kerr with that of an asymptotically
flat end with vanishing Simon tensor.
The first objection to this characterisation
is, precisely, that the isometry with Kerr is only established
in some neighbourhood of infinity, and hence the extension of this isometry
to the whole (strict) stationary region cannot be ensured yet.
This motivates, in fact, the search for improved local characterisations,
since the problem of the extension of the isometries 
to whole (strict) stationary regions
may be fixed by exploiting the
local characterisations to their full extent.
Indeed
Perj\'es found \cite{perjessimon}
that the most general metric with vanishing Simon tensor
depends only on a few parameters, and thus showed that the asymptotically
flatness condition in the characterization of Kerr is only
necessary in order to fix the value of some constants.
In this paper we thus follow an analogous aim, since we exhaust
the implications of the vanishing of the Cotton tensor
in the static electrovacuum problem.

The second drawback the characterisation of Kerr in \cite{walterKerr} faces
is that, by construction, it is not valid within the ergosphere.
To address this problem 
Mars \cite{MARCKerr1,MARCKerr2}
managed to improve that characterisation
and include the ergosphere
by constructing the so called Mars-Simon tensor, this time
relative to the spacetime. The Kerr characterisation in \cite{MARCKerr1}
(see also Theorem 1 in \cite{MARCKerr2})
is essentially local, since the vanishing of the Mars-Simon tensor
produces a family of vacuum solutions depending on two complex constants,
only to be fixed by some simple global consideration.
On the other hand, in \cite{MARCKerr2}, Mars provided a
characterisation with a much weaker local condition,
using more effectively the asymptotic flatness.
The work in \cite{MARCKerr1}
has been extended recently by Wong in \cite{wong09},
by providing a couple of extended characterisations for the Kerr-Newman family,
the first being purely local.

The main assumptions inherent
to the spacetime characterisations of the 
Kerr-Newman family
have two crucial direct implications.
The first is the 
degeneration of the Weyl tensor
(type $D$), 
and the second
is the existence of a functional
relationship of the gravitational and electromagnetic potentials
in the static case.
None of these restrictions are taken as assumptions in the present work.
Not imposing any restriction on the Petrov type is important in the
static case, as otherwise the Majumdar-Papapetrou class would
not be taken into cosideration. On the other hand,
the key result in the present paper that leads to the
complete solution of the conformastat electrovacuum problem
is precisely
that the aligment of the gradients of the potentials
is necessary.
In this sense, in the static case
the results found here generalise completely those in \cite{wong09}.
Furthermore, these results suggest that the known
local Kerr-Newman characterisations may be improved
by relaxing some of the requirements involved.


The vanishing of the Cotton tensor in the stationary vacuum problem
was dealt with in a series of three papers by Luk\'asz \emph{et al.}
in \cite{perjes1} and Perj\'es in \cite{perjes2,perjes3}
(see also \cite{perjes3p}).
They found the whole set of
conformastationary\footnote{Let us note that they
refer to conformastationary spacetimes simply as ``conformastat''.}
vacuum spacetimes. 
In a first paper
\cite{perjes1} the solutions possessing
a functional relationship between the real and imaginary parts of the
Ernst potential $\ernst$ were found to consist of
three bi-parametric families of solutions
generated from the three conformastat vacuum solutions (Class $A$)
by the Ehlers transformation.
In \cite{perjes2}, using the 
purposely defined ``Ernst coordinates'', Perj\'es
found that solutions with functionally independent
real and imaginary parts of $\ernst$ necessarily admit
a spacelike isometry\footnote{The authors talk of an ``axial'' symmetry,
but no global property is involved in the result at this point.},
to conclude in \cite{perjes3} (see also \cite{perjes3p})
that this set of solutions is empty.
Therefore, all conformastationary vacuum spacetimes belong to the
three families presented in \cite{perjes1},
which can be thought as the NUT-type extensions
of Schwarzschild and its plane and hyperbolic counterparts.

The plan of this paper is analogous.
We start in  Sections \ref{sec:conformastationary} and
\ref{sec:commonframe}
by 
showing how the conformastat electrocavuum problem and the conformastationary
vacuum problem can be treated within a common framework by using a suitable
notation.
The motivation is to use
the previous works \cite{perjes2,perjes3,perjes3p} as a guide,
and additionally, to recover those results.
In Section \ref{sec:Lnotzero} we prove the key result: the
conformastat electrovacuum spacetimes necessarily contain a
functional relationship between
the gravitational and electromagnetic potentials.
Regarding the use of the procedures in references \cite{perjes2,perjes3}
two points
must be stressed. Firstly, the ``common'' proof needs at many stages a
different approach, since the variables involved in the general case
are not necessarily complex,
and thus the positivity of some products cannot be used.
Secondly, the final stages in the proof differ from those
in \cite{perjes3} and fix some errors found in \cite{perjes3p}.
Anyway, to ease the comparison with these works
we have kept the same notation whenever possible.

In the second part, Section \ref{sec:classLzero} is devoted to
complete the study of conformastat electrovacuum
spacetimes
by classifying and exploiting the necessary functional relationship
between the gravitational and electromagnetic potentials.
We find that \emph{all conformastat electrovacuum spacetimes either belong
to the Majumdar-Papapetrou class or correspond to
either the Bertotti-Robinson solution
or the exterior Reissner-Nordstr\"om solution toghether
with its plane and hyperbolic counterparts.}
Furthermore, the procedure used for
the classification provides an improved characterisation of
the Majumdar-Papapetrou class. This is known to be
the class of static electrovacuum spacetimes
such that the usual rescaled induced metric in the space of orbits is flat.
Here we find that one only needs to ask that metric to be
conformally flat and with vanishing Ricci scalar.

The main result constitutes then a completely local characterisation
of the static and charged (multi) black hole solutions,
plus the ``non-standard'' Majumdar-Papapetrou solutions,
the near-horizon geometry (Bertotti-Robinson)
and the plane and hyperbolic counterparts of the exterior
Reissner-Nordstr\"om.
A simple global consideration
can be used now to single out the black hole solutions.
The essentially local characterisation is thus that
\emph{the conformastat electrovacuum asymptotically flat spacetimes are
isometric either to the asymptotically flat subset of
the Majumdar-Papapetrou class or the Reissner-Nordstr\"om static exterior.}

Sign conventions differ from those in \cite{perjes1,perjes2,perjes3}
and follow those in \cite{sols}: the metric has signature $(-,+,+,+)$
and the Riemann tensor is defined so that
$2\nabla_{[\alpha}\nabla_{\beta]}w^\lambda
=R^\lambda{}_{\gamma\alpha\beta}w^\gamma$.
Greek indices refer to the spacetime and Latin indices to the
three-dimensional space of orbits.
Units are chosen so that $G = c = 1$.

\section{Conformastationary spacetimes}
\label{sec:conformastationary}
A stationary spacetime $\spacetime$
is \emph{locally} defined by the existence of a timelike
Killing vector field $\xi^\mu$, whose space of orbits
invariantly determines a
differentiable 3-dimensional Riemannian manifold $\Sigma_3$.
Local coordinates $\{t,x^a\}$ exist for which $\xi^\mu=\partial_t$
and such that the line-element can be cast as \cite{sols}
\begin{equation}
\label{eq:stationary}
ds^2=-e^{2U}(dt+ A_a dx^a)^2 + e^{-2U}\hh_{ab} dx^a dx^b,
\end{equation}
where $U$, $A_a$ and $\hh_{ab}$ do not depend on $t$. Applying
the usual projection formalism \cite{geroch71,sols}
we will think of $U$ as a function on $\Sigma_3$,
$A_a$ as a 1-form belonging to $T^*\Sigma_3$ and $\hh_{ab}$ as a metric on
$\Sigma_3$. 
Once these three objects are given, the local geometry of the stationary
spacetime $\spacetime$ is fully specified by using
(\ref{eq:stationary}).
Let us, from now on, endow $\Sigma_3$ with the metric $\hh_{ab}$
and use the first latin indices $a,b,\ldots$ for objects defined on
$\orbits$.
A \emph{conformastationary} spacetime is a stationary spacetime whose
space of orbits $\orbits$ is conformally flat \cite{sols}.
Thence, in a conformastationary spacetime there exist coordinates $\{x,y,z\}$
in which $\hh_{ab}dx^a dx^b=e^{\lambda(x,y,z)}(dx^2+dy^2+dz^2)$.
The intrinsic characterisation of a conformally flat
3-space is the vanishing of the Cotton tensor $C_{abc}$
\cite{sols,schouten}, or equivalently, the York tensor density \cite{york},
defined as $Y_a{}^e\equiv\hheta^{bce}C_{abc}$, where $\hheta_{abc}$
denotes the volume form of $(\Sigma_3,\hh_{ab})$,
which satisfies $Y_{ae}=Y_{ea}$ and $Y_a{}^a=0$.
More expliclitly,
$\Sigma_3$ is conformally flat if and only if
\begin{equation}
\label{eq:york}
Y_a{}^e=
\hheta^{bce}\left(2\hhd_c\RR_{ba}-\frac{1}{2}\hh_{ab}\hhd_c\RR\right)=0,
\end{equation}
where $\RR_{ab}$ and $\hhd$ denote the Ricci tensor and
covariant derivative relative to $\hh_{ab}$.

\emph{Conformastat} spacetimes are those conformastationary spacetimes
which are, in fact, static.
In this context, a static spacetime is thus characterised by $A_a=0$.


\subsection{Electrovacuum field equations}
Let us first fix one basic assumption and some notation.
First, we will restrict ourselves to Maxwell fields $F_{\alpha\beta}$
in $\spacetime$ which inherit the stationary symmetry,
i.e. for which $\mathcal{L}_\xi \bm F=0$.
The Einstein-Maxwell equations outside the sources imply (locally, at least)
the existence of two complex scalars, $\Phi(x^a)$ the electromagnetic
potential, and $\ernst(x^a)$ the Ernst potential.
These two potentials in $\orbits $
satisfy the so-called Ernst-Maxwell equations,
\begin{eqnarray}
\label{eq:divH}
\hhd^a H_a+\frac{1}{2}\, \cxc G\cdot H-\frac{3}{2}\, G\cdot H=0,\\
\label{eq:divG}
\hhd^a G_a-\cxc H\cdot H-(G-\cxc G)\cdot G=0,
\end{eqnarray}
where $H_a\equiv (\cxRe \ernst+\Phi\cxc\Phi)^{-1/2}\Phi_{,a}$ and
$G_a\equiv 1/2 (\cxRe\ernst+\Phi\cxc\Phi)^{-1}(\ernst_{,a}
+2 \cxc\Phi\Phi_{,a})$ and
the dot denotes the scalar product.
It will be convenient for later to
note two identities that $H_a$ and $G_a$ satisfy:
$\rmd H=H\wedge\cxRe G$ and $\rmd G=G\wedge \cxc G+\cxc H\wedge H$.
These relations are, in fact, the integrability
conditions for the two potentials.

The rest of the 
Einstein-Maxwell equations without sources reduce to the following problem for
$\hh_{ab}$
\begin{equation}
  \label{eq:Ricci}
  \RR_{ab}=G_a \cxc G_b+\cxc G_a G_b-(H_a\cxc H_b+\cxc H_a H_b).
\end{equation}
Once $\hh_{ab}$ is known,
the geometry and electromagnetic field are recovered from the complex
potentials. 
The metric function $U$ and the 1-form $A_a$ are determined by
the relations
\[
e^{2U}=\cxRe \ernst+\Phi\cxc\Phi,\quad
(\rmd A)_{ab}
=2 e^{-4U}\hheta_{abc}\cxIm G^c,
\]
taking into account that the freedom in the determination of $A_a$
corresponds to a transformation of the time coordinate
of the form $t\to t+\chi(x^a)$ \cite{sols}.
The electromagnetic field, conveniently described by the
self dual 2-form $\ff_{\mu\nu}$
\[
\bm\ff\equiv \bm F + i * \bm F,
\]
where $*$ stands for the Hodge dual in $\spacetime$,
i.e. $*F_{\alpha\beta}=\frac{1}{2}F^{\mu\nu}\eta_{\mu\nu\alpha\beta}$,
is thus recovered by
\[
\bm{\ff}=-e^{-U}\left[\bm H\wedge\bm \xi
  + i * (\bm H\wedge\bm \xi)\right],
\]
where the 1-form $H_\mu$ in $\spacetime$ is given, in coordinates
adapted to the Killing (\ref{eq:stationary}),
by $H_\mu=(0,H_a)$. Note that $\xi_\mu=-e^{2U}(1,A_a)$.
The real and imaginary parts of $H_\mu$ 
correspond to the electric and magnetic fields with respect
to the observer defined by $\bm u\equiv e^{-U}\bm \xi$,
this is $H_\mu=E_\mu+i B_\mu=\ff_{\mu\nu} u^\nu.$
For completeness, let us note that
the intrinsic definition of $G_\mu$ in $\spacetime$ is given by
$
 G_\mu\equiv u^\nu(\nabla_\nu u_\mu+ i * \nabla_\nu u_\mu)$
and its real and imaginary parts  $G_\mu=a_\mu+i\frac{1}{2} w_\mu$
correspond to the acceleration and twist vectors
of the congruence $\bm u$.

\subsection{Vacuum and electro-magnetostatic cases}
The stationary \emph{vacuum case} is characterised by $\Phi=0$, so that $H_a=0$
and hence (\ref{eq:Ricci}) specialises to 
\[
\RR_{ab}=G_a \cxc G_b+\cxc G_a G_b
\]
and the Ernst-Maxwell equations reduce to
\[
\hhd^a G_a-(G-\cxc G)\cdot G=0
.
\]
The integrability condition that $G_a$ satisfies
reads simply $\rmd G=G\wedge \cxc G$.

The \textit{static case} is characterised by $G_a-\cxc G_a(=2 \cxIm G_a)=0$.
Well known fact is that the conditions for $H_a$ and $G_a$ 
and the field equations yield $\rmd G=0$
(in fact $G_a=U_{,a}$),
and $\cxc H_a=e^{-2i\theta} H_a$ for some constant $\theta$ (see e.g. \cite{Das79}).
Let us now define the vector
$\eee_a\equiv e^{-i\theta} H_a$, which is real by construction
and related to the electric and magnetic static fields by
$E_a=\cos\theta \eee_a$ and $B_a=\sin\theta \eee_a$.
Instead of working with the complex $\Phi$
let us consider the real potential
$\Psi=e^{-i\theta}\Phi$, so that $\eee_a=e^{-U}\Psi_{,a}$.
We are thus left with two real vectors: $G_a$ and $X_a$.

The stationary vacuum and the static electrovacuum cases are known
to have an analogous structure,
although they are inequivalent (see e.g. Chapter 34 in \cite{sols}).
The analogy has been used previously in the literature
in a more or less implicit manner (see e.g. \cite{taub-bolt}).
The fact that the two problems
are inequivalent comes most notably from
the signature of the potential spaces,
which differ in the two cases.
Despite this,
one can make the analogy explicit,
and useful in the present study,
incorporating that change of signature by making use of a hyperbolic-complex
or motor number construction, based on the
real Clifford algebra $C\ell_{1,0}(R)$
(see e.g. \cite{lambertpiette} and references therein),
for the static electrovacuum problem.
We call $j$ the hyperbolic imaginary unit, which satisfies $j^2=1$,
and denote the conjugate operation by $\jc j=-j$.
Note that $C\ell_{0,1}(R)$ is isomorphic to the field of complex numbers,
in which $i$ is the elliptic imaginary unit.

We are now ready to define
\[
\sss_a\equiv \frac{1}{2}(G_a+j\eee_a),
\]
in terms of which the Eintein-Maxwell and Ernst-Maxwell equations
read
\begin{eqnarray}
  \label{eq:ricciS}
  \RR_{ab}=4(\sss_a\jc\sss_b+\jc\sss_a \sss_b),\\
  \label{eq:divS}
  \hhd^a \sss_a-(\sss-\jc \sss)\cdot \sss=0,
\end{eqnarray}
and the identities for $G_a$ and $X_a$ reduce to
$\rmd \sss=\sss\wedge \jc \sss$.


\section{A common framework}
\label{sec:commonframe}
For the sake of completeness and to allow us to use
the techniques and some results of previous works on
conformastationary vacuum spacetimes \cite{perjes1,perjes2},
we set up a common and more general problem using a common notation.

Let us denote by $\genj$ any of both the complex $i$ and the hypercomplex $j$,
so that $\genj^2=\pm 1$ accordingly, and the
general conjugation by $\,\,\opp{ }\,\,$, so that $\opp \genj=-\genj$
stands for either $\cxc i=-i$ or $\jc j=-j$.
Any object of the form $F=f+\genj g$ will be called a composed
object, and 
$\oppRe(F)$ and $\oppIm(F)$ will denote its real and imaginary parts.

Consider now a composed vector field $\yy^a$ and a real metric $\hh_{ab}$
which satisfy the system of equations
\begin{eqnarray}
  \label{eq:ricciY}
  \RR_{ab}=\nn(\yy_a\opp\yy_b+\opp\yy_a \yy_b),\\
    \label{eq:divY}
  \hhd^a \yy_a-(\yy-\opp \yy)\cdot \yy=0,\\
  \label{eq:dY}
\rmd \yy=\yy\wedge \opp \yy.
\end{eqnarray}
One could regard this problem at the level of the potentials, but for our
purposes it suffices to set up the problem for the vectors
and thence include the integrability conditions as equations.
The vacuum case is recovered by taking $\nn=1$, $\yy_a=G_a$, a complex 1-form,
and the conjugate being the complex conjugate. The static case
corresponds to $\nn=4$, $\yy_a=\sss_a$, a $j$-1-form and the conjugate
being the $j$-conjugation. Note that in both cases the right-hand side
of the equation for $\RR_{ab}$ is, as it should, a real quantity,
whereas the equations (\ref{eq:divY}) and (\ref{eq:dY})
yield two real equations each.

\subsection{Conformastationarity}
Conformastationarity follows by
the vanishing of the York tensor density of $\hh_{ab}$, this is,
by applying equation (\ref{eq:york}) to the Ricci tensor
as expressed in (\ref{eq:ricciY}).
Before writing down the explicit expressions, let us introduce
a very convenient vector (see \cite{perjes1})
\footnote{Although we have kept
the notation as close as possible to that used in \cite{perjes1},
the vector $\lll$ defined here differs by a multiplicative $i$.}
\[
\lll\equiv \star(\yy\wedge\opp\yy),
\]
where $\star$ denotes the Hodge-dual in $\orbits$,
i.e. $\lll_a=\yy^b\opp\yy^c\hheta_{bca}$.
By construction we have $\opp \lll_a=-\lll_a$ 
and $\lll\cdot\yy=\lll\cdot\opp\yy=0$. Note also
that $\lll=\star\rmd \yy$.
Let us stress the fact that since $\opp \lll_a=-\lll_a$,
$L_a$ is imaginary and thus $\genj^2 ( \lll\cdot \lll)\geq0$.
Introducing (\ref{eq:ricciY}) into (\ref{eq:york}) one obtains
the real equation
\begin{equation}
  \label{eq:yorky}
  \fl
  \frac{1}{2\nn} Y_a{}^{e}=(\yy_a-\opp\yy_a)\lll^e+
  \hheta^{bce}(\opp\yy_b\hhd_c\yy_a+\yy_b\hhd_c\opp\yy_a)
  -\frac{1}{2}\hh_{ab}\hheta^{bce}\hhd_c(\yy\cdot\opp\yy)=0.
\end{equation}
Since $Y_a{}^a=0$ this equation contains at most 5 independent components.
We will exploit the consequences of those equations later.

Two very different
situations arise in the study of the system of equations composed by
(\ref{eq:divY}), (\ref{eq:dY}), (\ref{eq:ricciY}) and (\ref{eq:yorky}),
for $\yy_a$ and $\hh_{ab}$:
the class of solutions for which $\lll_a\neq 0$ and those for which
$\lll_a=0$.
Nevertheless, before entering into the study of these two cases
one has to consider the case $\yy\cdot\yy=0$.
In the static case $\yy_a=\sss_a$ is $j$-composed and $\sss\cdot\sss=0$
implies, in particular, $G\cdot G+\eee\cdot\eee=0$,
which clearly leaves us only with the trivial case $G_a=\eee_a=0$.
However, in the vacuum case $\yy_a=G_a$ is complex and one can have, in principle,
fields for which $G\cdot G=0$. The study of these null fields
was performed in \cite{perjes1}, where it was proven that
no null coformastationary vacuum spacetimes exist apart
from the trivial case of flat spacetime.
In the following we will therefore take $\yy\cdot\yy\neq 0$
without loss of generality.

\section{The class $\lll_a\neq  0$} 
\label{sec:Lnotzero}
In this section we prove that the class $\lll_a\neq  0$ is
empty in two steps. We first show that if $\lll_a\neq  0$
there must be an additional isometry, and then that the existence
of that isometry implies the non-existence of solutions with $\lll_a\neq  0$.

Let us take the basis $\{\lll_a,\yy_a,\opp\yy_a\}$. (Note that the
associated basis for the real tangent vector space is
composed by $\genj L_a$, $\yy_a+\opp\yy_a$ and $\genj (\yy_a-\opp\yy_a)$.)
The metric $\hh_{ab}$ expressed in this basis reads
\begin{equation}
\label{eq:hhY}
\fl
\hh_{ab}=\frac{1}{\lll\cdot \lll}
\left[
\lll_a \lll_b
+(\yy\cdot\yy)\opp\yy_a\opp\yy_b
+(\opp\yy\cdot\opp\yy)\yy_a\yy_b
-(\yy\cdot\opp\yy)(\yy_a\opp\yy_b+\opp\yy_a\yy_b)
\right],
\end{equation}
where $\lll\cdot\lll=(\yy\cdot\yy)(\opp\yy\cdot\opp\yy)-(\yy\cdot\opp\yy)^2$.
Since $\lll_a\neq  0$ we have $\genj^2 (\lll\cdot\lll)>0$.


Using the obvious notation
$Y_a{}^b\yy_b$ by $Y_a{}^\yy$, etc..., the tracefree property
of $Y_a{}^e$ translates onto
\[
Y_\lll{}^\lll=
-(\yy\cdot\yy)Y_{\opp\yy}{}^{\opp\yy}
-(\opp\yy\cdot\opp\yy)Y_{\yy}{}^{\yy}
+(\yy\cdot\opp\yy)(Y_{\yy}{}^{\opp\yy}+Y_{\opp\yy}{}^{\yy}).
\]
Together with the use of the conjugate operation, this allows
us to keep all the information contained in $Y_a{}^b$
in only three components: $Y_\yy{}^\yy$, $Y_{\opp\yy}{}^\yy$
and $Y_\lll{}^{\yy}$. (Note that $\opp{Y_{\opp\yy}{}^\yy}=Y_{\yy}{}^{\opp\yy}$.)
The corresponding three equations in (\ref{eq:yorky}),
from where the five real independent equations eventually follow, read
\begin{eqnarray}
  \label{eq:Yyy}
  \lll^c\hhd_c(\yy\cdot\yy)=0,\\
  \label{eq:Yycy}
  \lll\wedge\rmd\lll=0,\\
  \label{eq:Yyl}
  \lll^c\lll^a\hhd_c\yy_a
  -\frac{1}{2}\hheta^{bce}\yy_e\lll_b\hhd_c(\yy\cdot\opp\yy)=0.
\end{eqnarray}

The interpretation of the equations (\ref{eq:Yyy}) and (\ref{eq:Yycy})
is straightforward.
Equation (\ref{eq:Yycy}) states that $L_a$ is hypersurface orthogonal,
i.e. integrable. 
Equation (\ref{eq:Yyy}) 
implies that the product $\yy\cdot\yy$ 
is constant along $\lll_a$.
In the static case this translates to the fact that
the two scalars $G^2+\eee^2$ and $G\cdot\eee$ are constant along the
direction orthogonal to the planes spanned by $G_a$ and $\eee_a$.

\subsection{The additional isometry}
\label{sec:addsymm}
In this subsection (together with \ref{appendixA})
we prove that the above equations
(\ref{eq:Yyy}), (\ref{eq:Yycy}) and (\ref{eq:Yyl}),
together with the Ricci equations (\ref{eq:ricciY})
and the integrability condition (\ref{eq:dY}) imply the existence
of a further isometry along $\lll_a$.


Since $\lll_a$ is integrable (\ref{eq:Yycy}) and imaginary,
there exist two real functions
$\chi(x^a)$ and $\varphi(x^a)$ such that
\begin{equation}
\label{eq:Lphi}
\lll=\genj\,\chi\rmd \varphi.
\end{equation}
The function $\varphi$ cannot be constant precisely because $\lll_a\neq0$,
and we can also take $\chi>0$ without loss of generality.
The integrability equations (\ref{eq:dY}) imply, in turn, the existence of two
further real functions, encoded in the composed potential $\sigma(x^a)$
so that\footnote{A simple inspection shows the relationship of
$\sigma$ with the original potentials. In the static case one has
$
\sigma=\frac{1}{2}(e^{U}+j\Psi),
$
whereas in the vacuum case one recovers the usual Ernst complex potential
in vacuum
$
\sigma(=\ernst)=e^{2U}+i\Omega.
$
}
\begin{equation}
\label{eq:yys}
\yy=\frac{1}{\sigma+\opp\sigma}\rmd \sigma.
\end{equation}

The main idea  is to use the three potentials
$\sigma$, $\opp\sigma$ and $\varphi$, as coordinates.
In the vacuum (complex) case \cite{perjes2} these particularise
to the so-called Ernst coordinates.
The independence of $\varphi$ and $\sigma$
is ensured by the orthogonality of $\lll_a$ and $\yy_a$.
Let us label this coordinate system as
\[
x^1=\sigma,\quad
x^2=\opp\sigma,\quad
x^3=\varphi.
\]
The real coordinates and manifold related quantities can always be
recovered by the obvious linear transformations to the coordinates
$\sigma+\opp\sigma$ and $\genj(\sigma-\opp\sigma)$.
There exists a freedom in choosing $\varphi$, since $\lll_a$ is invariant
under the transformation
\begin{equation}
\label{eq:x3free}
\varphi\to f(\varphi),\quad
\chi\to\chi\left(\frac{df}{d\varphi}\right)^{-1},
\end{equation}
for any smooth function $f$ with non-vanishing derivative.
This freedom will be
only used in the last step of the proof (see \ref{appendixA}).

The form of the metric $\hh_{ab}$ in these coordinates follows
directly from (\ref{eq:hhY}) together with
(\ref{eq:Lphi}) and (\ref{eq:yys}). With the help of a shorter notation
for the products
\begin{eqnarray*}
\alpha\equiv (\sigma+\opp\sigma)^2(\yy\cdot\yy),\quad 
\gamma\equiv \opp\alpha=(\sigma+\opp\sigma)^2(\opp\yy\cdot\opp\yy),\\
\beta\equiv (\sigma+\opp\sigma)^2(\yy\cdot\opp\yy),
\quad
\end{eqnarray*}
(where note that $\alpha$
is composed, we denote by $\gamma$ its conjugate 
and $\beta=\opp\beta$ is real)
together with the auxiliary real functions $\rho$,
which essentially substitutes $\chi$, and $D$ defined by
\begin{equation}
\fl
\label{eq:rhoD}
D\equiv \alpha\gamma-\beta^2=(\sigma+\opp\sigma)^4
(\lll\cdot\lll),\quad
\rho^2\equiv\frac{\genj^2\chi^2}{\lll\cdot\lll}
=\genj^2 D^{-1}(\sigma+\opp\sigma)^4\chi^2>0,
\end{equation}
the line-element reads
\begin{equation}
\label{eq:hh123}
\hh_{ab} dx^a dx^b=\frac{1}{D}
\left[\gamma (dx^1)^2 -2\beta dx^1dx^2+\alpha (dx^2)^2
+\rho^2 D (dx^3)^2\right].
\end{equation}
Since we are dealing with $\yy\cdot\yy\neq 0$, $\alpha$ cannot
vanish, and
in the complex case one thus readily has that $\alpha\gamma>0$
because $\alpha\gamma=\alpha\cxc\alpha$.
But in the $j$-composed case this is not ensured a priori.
Nevertheless, the real function $D$ satisfies $\genj^2 D>0$ by construction,
which in the $j$-composed case translates onto  $D>0$
and therefore $\alpha\gamma>0$ necessarily.
To sum up, \emph{in any case we have} $$\alpha\gamma>0.$$
Let us also remark
that $\det\hh=\rho^2 D^{-1}$ in these composed coordinates.
On the other hand, given
(\ref{eq:Lphi}) and (\ref{eq:yys}) together with the definition of $\lll_a$,
the volume element is fixed by
$\hheta_{123}=\genj \rho D^{-1}\sqrt{\genj^2 D}$.
We will take the metric
to be determined by the four real unknown functions encoded in
$\alpha$, $\beta$ and $\rho$. Without loss of generality
we take $\rho>0$.

It only remains to write equations (\ref{eq:divY}),
(\ref{eq:ricciY}) plus (\ref{eq:Yyy}) and (\ref{eq:Yyl}) in this coordinate
system.
Since
\[
\vec\lll=\genj\frac{\sqrt{\genj^2 D}}{\rho(\sigma+\opp\sigma)^2}\partial_{3},
\]
equation (\ref{eq:Yyy}) holds iff $\alpha$ and
$\gamma$ are functions of $\sigma$ and $\opp\sigma$ only.
With this information at hand equation (\ref{eq:divY})
translates onto
\begin{equation}
\label{eq:1}
(\alpha\partial_1 + \beta \partial_2)\ln\frac{\rho}{\sqrt{\genj^2 D}}
+\partial_1\alpha+\partial_2 \beta=\frac{2\alpha}{\sigma+\opp\sigma},
\end{equation}
while (\ref{eq:Yyl}) reads
\begin{equation}
  \label{eq:2}
  2(\alpha\partial_1 + \beta\partial_2)\ln\rho+\partial_2\beta
-\frac{2\beta}{\sigma+\opp\sigma}=0.
\end{equation}
The components of the equation for the Ricci tensor (\ref{eq:ricciY})
yield the four independent composed equations
\begin{equation}
  \label{eq:ricci123}
  \RR_{11}=\RR_{13}=\RR_{33}=0,\quad
  \RR_{12}=\frac{\nn}{(\sigma+\opp\sigma)^2},
\end{equation}
which encode the six real equations, due to the fact that
$\RR_{22}=\opp{\RR_{11}}$, $\RR_{23}=\opp{\RR_{13}}$.
Note that $\RR_{12}$ and $\RR_{33}$ are real.

The first consequence the integrability conditions
of the system (\ref{eq:1})-(\ref{eq:ricci123}) provide is the following:

\begin{prop}
\label{res:beta3}
Any solution of the system of
equations (\ref{eq:1}),(\ref{eq:2}) and (\ref{eq:ricci123})
for $\iota^2 D>0$ necessarily has $\beta_{3}(\equiv\partial_3\beta)=0$.
\end{prop}
To ease the reading the proof is left to \ref{appendixA}.
\fin

The only remaining function in the line-element
(\ref{eq:hh123}) which may still depend
on $x^3$ is $\rho$. But this cannot be the case due to
(\ref{eq:2}). 
Assuming that a solution to (\ref{eq:2}) exists,
integration of $\gamma(\ref{eq:2})-\beta\opp{(\ref{eq:2})}$ yields
\[
\ln \rho=\int^{\sigma}_0\frac{1}{2D}\left(\beta\beta_1-\gamma\beta_2
+2\beta\frac{\gamma-\beta}{x^1+\opp\sigma}dx^1\right) + \ln\rho_0.
\]
The integral does not depend on $x^3$, but the arbitrary term $\rho_0$
depends on $x^3$, in principle. However, this term can be eliminated
by using remaining freedom in choosing the
coordinates (\ref{eq:x3free}),
given by a transformation $x^{3'}=x^{3'}(x^3)$.

This completes the proof of the exitence of an additional
spacelike isometry whenever $\opp{\yy}\wedge\yy\neq 0$ and
$\yy\cdot\yy\neq 0$.


\subsection{The class $\lll_a\neq0 $ is empty}
\label{sec:Lnozeroempty}
In \ref{appendixB} we prove the following result:
\begin{prop}
\label{teo:noL}
There is no solution of the system of equations
(\ref{eq:1}), (\ref{eq:2}) and (\ref{eq:ricci123})
for functions depending on $\sigma$ and $\opp\sigma$ only, with
$\iota^2 D>0$, for $N\neq 16,-2,8/5$.\finn
\end{prop}
Since we are interested only in the cases $N=1$ and $N=4$ we do not
investigate further the compatibility of (\ref{eq:1}), (\ref{eq:2})
and (\ref{eq:ricci123})
for the special cases $N= 16,-2,8/5$.

This proposition thus states that the class $\lll_a\neq 0$ with and additional
isometry is empty. Combined with Proposition \ref{res:beta3}
this finally implies that the full class $\lll_a\neq 0$ is empty.

We are thus only left with $\lll_a=0$ necessarily.
This means that $\yy_a$ and $\opp\yy_a$ are parallel,
and $\rmd \yy=0$ by (\ref{eq:dY}).
Therefore
$\yy_a$ is a gradient of some composed potential
whose real and imaginary parts are functionally dependent.

In particular, on the one hand we have thus recovered the result
found in the series of papers
\cite{perjes1,perjes2,perjes3} (see also \cite{perjes3p}):
\begin{theorem}
Conformastationary vacuum spacetimes are always characterised
by a functional relation between the potentials $U$ and $\Omega$.\finn
\end{theorem}
On the other hand, in the stationary electrovacuum case we have thus proven: 
\begin{theorem}
Conformastat electrovacuum spacetimes are always characterised
by a functional relation between the potentials $U$ and $\Phi$.\fin
\end{theorem}

\section{The complete solution of the conformastat electrovacuum
problem}
\label{sec:classLzero}
In the conformastationary vacuum case
the complete solution is thus given by those spacetimes for which
$U=U(\Omega)$.
This was studied in \cite{perjes1}. The solution consists of three
explicit bi-parametric families of line-elements, as described in the
Introduction.
We refer to \cite{perjes1} for the explicit form of the line-elements.

In the following we focus on the the static case.
From the above results we know we only have to look for solutions
for which
\[
U=U(\Psi).
\]
This is a well known ansatz used to find electro(-magneto)static
solutions as described in \cite{sols}, Section 18.6.3.
Our work consists on finding \emph{all} the conformastat
solutions among this class.

The divergence equation (\ref{eq:divY})
firstly fixes the functional relationship
to be\footnote{The relationship one obtains is in fact
$e^{2U} = b +a \Psi + \Psi^2$ for arbitrary constants $a$ and $b$.
The constant $b$ can be rescaled by
using the freedom $\Psi\to \Psi+const.$ (if $b\leq 0$) or
a rescaling of the $t$ coordinate (if $b>0$).} (see e.g. \cite{sols})
\[
e^{2U} = 1 - 2 c \Psi + \Psi^2,
\]
for an arbitrary constant $c$, which can be rewritten in parametric form
in terms of an auxiliary function $V$ as
\begin{eqnarray}
\fl
c^2 = 1: \qquad &\Psi = c - 1/V, \qquad \qquad &e^{2U} = V^{-2}, \label{eq:k1} \\
\fl
c^2 > 1: &\Psi = c - \sqrt{c^2 - 1} \coth V, \qquad &e^{2U} = (c^2 - 1) \sinh^{-2} V,
\label{eq:k2} \\
\fl
c^2 < 1: &\Psi = c - \sqrt{1 - c^2} \cot V, \ &e^{2U} = (1 - c^2) \sin^{-2} V, \label{eq:k3}
\end{eqnarray}
and secondly implies $$\hhd^2 V=0$$ in all cases.
The Ricci equations (\ref{eq:ricciY})
reduce now to
\begin{eqnarray}
\fl
  c^2 = 1: \qquad&\RR_{ab} = 0,\label{eq:riccif_flat}\\
\fl
  c^2 > 1: &\RR_{ab} = 2 V_{,a} V_{,b}, \qquad\\
\fl
  c^2 < 1: &\RR_{ab} = - 2 V_{,a} V_{,b}.
\end{eqnarray}
The remaining equation that $\hh_{ab}$ and $V_a$ have
to satisfy corresponds to the conformal flatness of $\hh_{ab}$,
and is encoded in (\ref{eq:yorky}).

Let us stress the fact that either case $c^2>1$ or $c^2<1$
constitutes a more general problem for $\hh_{ab}$
than the problem for the conformally flat 3-metric
one encounters in the black hole (global) uniqueness theorems
(see e.g. \cite{heusler}).
In the uniqueness theorems for charged black holes
one establishes from global considerations (using the positive mass theorem)
not only the conformal flatness
of the 3-metric and that the potentials are functionally related,
but also that the conformal factor depends only on the potential.
Since the conformal factor is not fixed a priori in the present study,
we cannot use the usual results found in the uniqueness theorems.
Instead we follow the procedure used by Das in \cite{Das71}
in the obtaining of the static vacuum solutions.

\subsection{Case $c^2=1$}
Equation (\ref{eq:riccif_flat})
does not involve $V_a$ and simply implies that $\hh_{ab}$ must be flat,
which in turn renders (\ref{eq:yorky}) to be automatically satisfied.
This is the well known Majumdar-Papapetrou class of solutions
\cite{sols}.
Given any solution
$V$ of the Laplace equation $\hhd^2V=0$ in flat 3-space,
the metric of the corresponding member
of the Majumdar-Papapetrou class
is found by using (\ref{eq:k1}), and thus reads
\[
    ds^2=-\frac{1}{V^2} dt^2 + V^2(dx^2+dy^2+dz^2),
\]
while the electromagnetic potential
$\Phi=e^{i\theta}\Psi$, after a trivial shift, is given by
\[
\Phi=-e^{i\theta}\frac{1}{V}.
\]

\subsection{Case $c^2 > 1$}
In this case we are looking for solutions
$\{\hh_{ab},V\}$ with $V_a\equiv V_{,a}\neq 0$ of the system
\begin{eqnarray}
\RR_{ab}=2V_{a}V_{b},\label{eq:1ricciV}\\
\hhd^2V=0,\label{eq:laplace}\\
4 V_{[b}\hhd_{c]} V_a-\hh_{a[b}\hhd_{c]}(V\cdot V)=0,\label{eq:yorkV}
\end{eqnarray}
where the latter stands for (\ref{eq:york}).

Because of $\hhd^2V=0$ and $V_{,a}\neq 0$, local coordinates $\{x,y^A\}$
with $A=2,3$ can be chosen so that $V=x$, and also
such that $\{y^A\}$ span the surfaces $S_2$ orthogonal
to $V_a$.
In these coordinates adapted to $V_a$
equation (\ref{eq:yorkV}) 
implies the following form of the metric
\begin{equation}
  \label{eq:ds2V}
  \hh_{ab}dx^a dx^b=W^2(x) dx^2+W(x) \Omega_{AB}dy^A dy^B, 
\end{equation}
where $W(x)$ is an arbitrary positive $C^3$ function and $\Omega_{AB}$
is a Riemannian $C^3$ metric on $S_2$, depending only on $\{y^A\}$.
The imposition of (\ref{eq:1ricciV}) leads to an equation for
$W(x)$ whose solution reads
\begin{equation}
  \label{eq:1W}
  W=(Ae^x+Be^{-x})^{-2}
\end{equation}
with constants $A$ and $B$.

It only remains to see that the surfaces $(S_2,\Omega_{AB})$ are of constant
curvature.
Let us consider the unit normal to $S_2$,
$n_a=W V_a$, and two vectors tangent to $S_2$, $e_{A}{}^a$,
this is $n_a e_{A}{}^a=0$, such that
$W\Omega_{AB}=e_{A}{}^a e_{B}{}^b \hh_{ab}$.
The second fundamental form of $S_2$ in $\Sigma_3$ thus reads
$K_{AB}=e_{A}{}^a e_{B}{}^b\nabla_a n_b=W'/(2W)\Omega_{AB}$.
On the other hand,
taking into account the identity between the Riemann and the Ricci tensors
in a 
3-dimensional space,
equation (\ref{eq:1ricciV})
is used to obtain the following expression of the Riemann
tensor of $\hh_{ab}$ projected on $S_2$
\[
\RR_{abcd}e_{A}{}^a e_{B}{}^b e_{C}{}^c e_{D}{}^d=
\Omega_{AC}\Omega_{BD}-\Omega_{AD}\Omega_{BC}.
\]
This expression is then introduced into the Gauss equation
in order to obtain the Riemann tensor for
$W\Omega_{AB}$ on $S_2$,
\begin{eqnarray*}
{}^{(W\Omega)}R_{ABCD}&=&\RR_{abcd}e_{A}{}^a e_{B}{}^b e_{C}{}^c e_{D}{}^d
+K_{AC}K_{BD}-K_{AD}K_{BC}=\\
&=&\left(\frac{W'^2}{4W^2}-1\right)(\Omega_{AC}\Omega_{BD}-\Omega_{AD}\Omega_{BC})\\
&=&-4AB W(\Omega_{AC}\Omega_{BD}-\Omega_{AD}\Omega_{BC}).
\end{eqnarray*}
The Riemann tensor for $\Omega_{AB}$ on $S_2$ thus reads
\[
{}^{(\Omega)}R_{ABCD}=W^{-1}{}^{(W\Omega)}R_{ABCD}
=-4AB (\Omega_{AC}\Omega_{BD}-\Omega_{AD}\Omega_{BC}).
\]
Therefore $(S_2,\Omega_{AB})$ is a surface of constant curvature $-4AB$.
In principle,
three different possibilities arise: (i) $AB=0$, (ii) $AB<0$ and (iii) $AB>0$.

\subsubsection{Case (i)}
This case is characterised by a flat $\Omega_{AB}$.
Coordinates $\{\vartheta,\varphi\}$ can therefore be chosen such that
\[
\Omega_{AB}dx^A dx^B=
d\vartheta^2+d\varphi^2.
\]
By changing
$x\to -x$ if necessary, we can take
$B=0$ and $A\neq 0$ without loss of generality,
so that $W=A^{-2} e^{-2x}$.
The line-element and electromagnetic potential are now
obtained by introducing this into (\ref{eq:ds2V}) and using (\ref{eq:k2}).
By performing the change $x=\frac{1}{2}\ln(b/(r+b))$ with
$b \equiv(4A^2\sqrt{c^2-1})^{-1}>0$,
together with $\tau=2\sqrt{c^2-1}\, t$, which induces the rescaling
$\Psi\to\Psi/(2\sqrt{c^2-1})$, the line-element can be finally cast as
\begin{equation}
  \label{eq:ds2i}
  ds^2=-\frac{(r+b)b}{r^2}d\tau^2
  +\frac{r^2}{(r+b)b}dr^2+r^2(d\vartheta^2+d\varphi^2),
\end{equation}
after a further convenient rescaling of $\vartheta,\varphi$.
The electromagnetic potential, after a trivial shift, reads
\[
\Phi=e^{i\theta}\frac{b}{r}.
\]
Note that the only restriction of the ranges of the coordinates
is on $r$. Since we have taken $b>0$ we are left with two different
ranges, $-b<r<0$ and $r>0$.
This family of solutions belong to the static \emph{plane-symmetric}
Einstein-Maxwell fields for which the surface element of
the surfaces $S_2$ with metric $e^{-2U}W\Omega_{AB}$ has a non-vanishing gradient
(see Chapter 15.4  in \cite{sols}).
It can also be regarded as the flat counterpart of
the Reissner-Nordstr\"om metric.
Although that family of spacetimes in  \cite{sols}
presents, in principle, two parameters $m$ and $e$,
whenever $m\neq 0$ a convenient change in $r$
can bring both $e$ and $m$ in \cite{sols} into a single parameter.
If $m=0$ that family falls into the $\RR_{ab}=0$ case.

\subsubsection{Case (ii)}
This case is characterised by a $\Omega_{AB}$ with positive
constant curvature $-4AB=4|AB|$.
Coordinates $\{\vartheta,\varphi\}$ can therefore be chosen such that
\[
\Omega_{AB}dx^A dx^B=
\frac{1}{4|AB|}(d\vartheta^2+\sin^2\vartheta d\varphi^2),
\]
where $\vartheta\in(0,\pi)$ and $\varphi\in[0,2\pi)$. 
After the change $e^{2x}=R^2$ and renaming $a\equiv 2\sqrt{c^2-1}$,
the direct substitutions lead to the line-element
\begin{equation}
\label{eq:ii}
\fl
ds^2=-\frac{a^2 R^2}{(R^2-1)^2}dt^2
+\frac{(R^2-1)^2}{a^2(AR^2+B)^4}\left[dR^2+\frac{(AR^2+B)^2}{4|AB|}
(d\vartheta^2+\sin^2\vartheta d\varphi^2)\right],
\end{equation}
for an electromagnetic potential given by
\[
\Phi=e^{i\theta} \frac{a}{2}\frac{R^2+1}{R^2-1}.
\]
Note that although three parameters appear in the metric, one of them
can be absorved applying a convenient change of coordinates,
and therefore only two are relevant. 
Now, this metric contains two very different subfamilies, depending
on whether the gradient of the surface element of the
$\{\vartheta,\varphi\}$ surfaces (see above) 
vanishes or not. Direct computation shows that the
gradient vanishes if and only if $A+B=0$.

When $A+B\neq 0$ one must obtain the Reissner-Nordstr\"om solution.
Indeed, the change $\{t,x\}\to\{\tau,r\}$ given by
\begin{equation}
\label{eq:changecoor}
e^{2x}=\frac{1-4\sqrt{|AB|(c^2-1)}Br}{1+4\sqrt{|AB|(c^2-1)}Ar},\quad
\tau=2\epsilon \frac{\sqrt{|AB|(c^2-1)}}{A+B}t,
\end{equation}
where $\epsilon^2=1$,
followed by the rearranging of the constants $A,B$ into
\begin{equation}
Q_c\equiv\frac{\epsilon}{4|AB|\sqrt{c^2-1}},\quad
M\equiv\frac{B-A}{8|AB|^{3/2}\sqrt{c^2-1}}
\label{eq:MQ}
\end{equation}
leads to the Reissner-Nordstr\"om metric in
canonical coordinates
\begin{equation}
\label{eq:R-N}
\fl
ds^2=-\left(1-\frac{2M}{r}+\frac{Q_c^2}{r^2}\right) d\tau^2
+\left(1-\frac{2M}{r}+\frac{Q_c^2}{r^2}\right)^{-1}dr^2
+r^2(d\vartheta^2+\sin^2\vartheta d\varphi^2),
\end{equation}
in the ranges $0<r<M-\sqrt{M^2-Q_c^2}$ and $M+\sqrt{M^2-Q_c^2}<r$,
and its corresponding electromagnetic potential
\[
\Phi=e^{i\theta}\frac{Q_c}{r}
\]
after a trivial shift. Note that $M^2-Q_c^2>0$ by construction
(see below) and that the usual $Q$ and $P$ \cite{heusler}
obviously correspond to $\cos\theta Q_c$ and $\sin\theta Q_c$
respectively.

The line-element of the special family for which $A+B= 0$ can be
conveniently writen as
\begin{equation}
\label{eq:ber-rob}
ds^2=-\sinh^2\left(\frac{z}{b}\right)d\tau^2+dz^2
+b^2(d\vartheta^2+\sin^2\vartheta d\varphi^2)
\end{equation}
for
\[
\Phi=e^{i\theta} \cosh\left(\frac{z}{b}\right)
\]
after the changes $\tau=\sqrt{c^2-1}t$ and $\sinh x=[\sinh(y/b)]^{-1}$,
where $1/b\equiv 4A^2\sqrt{c^2-1}$.
This is the well known Bertotti-Robinson solution,
which is also characterised by being the
only homogeneous Einstein-Maxwell field with a
homogeneous non-null Maxwell field,
and the only conformally flat solution with a non-null
Maxwell field \cite{sols}.
Furthermore, the Bertotti-Robinson solution is known to describe
the \emph{near-horizon} limit of an extreme Reissner-Nordstr\"om
black hole \cite{carter72}.

It is worth noticing here that the relationship
$4|AB|(M^2-Q_c^2)=(A+B)^2Q_c^2$ implies that in this class (ii) of solutions
we are only finding the $M^2-Q_c^2>0$ part of the Reissner-Nordstr\"om
solution. Indeed, the extreme
case $M^2-Q_c^2=0$ is excluded in this class (ii) because $A+B=0$
in (\ref{eq:ii}) leads to
the Bertotti-Robinson solution instead.
This is due to the fact that
in this case (ii) we are considering solutions with $\RR>0$
whereas the extreme Reissner-Nordstr\"om solution has $\RR=0$,
thus falling into the Majumdar-Papapetrou class.
The $M^2-Q_c^2<0$ case implies $\RR<0$, and therefore will appear
in the case $c^2<1$ below.
To sum up, the line-element (\ref{eq:ii}) corresponds to the
(static and $M^2-Q_c^2>0$) Reissner-Nordstr\"om solution containing the
near-horizon Bertotti-Robinson metric as a limit instead of the extreme case.
Note, again, that only two parameters in (\ref{eq:ii}) are relevant,
but for the sake of shortness we do not pursue the rewritting of
(\ref{eq:ii}) any further.

\subsection{Case (iii)}
This case is characterised by a $\Omega_{AB}$ with negative
constant curvature $-4AB$.
Coordinates $\{\vartheta,\varphi\}$ can therefore be chosen such that
\[
\Omega_{AB}dx^A dx^B=
\frac{1}{4AB}(d\vartheta^2+\sinh^2\vartheta d\varphi^2),
\]
where $\vartheta\in(0,\infty)$ and $\varphi\in(-\infty,\infty)$. 
As in the previous case (ii),
after the change $e^{2x}=R^2$ and $a\equiv 2\sqrt{c^2-1}$,
the direct substitutions lead to the same line-element
(\ref{eq:ii}) 
with $\sin\vartheta$ changed by $\sinh\vartheta$.

Since $AB>0$ in this case, $A+B$ cannot vanish, and
therefore the change (\ref{eq:changecoor}) is always possible.
After performing the same parameter redefinitions (\ref{eq:MQ})
one obtains the metric
\begin{equation}
\label{eq:pseuR-N}
\fl
ds^2=-\left(-1-\frac{2M}{r}+\frac{Q_c^2}{r^2}\right) d\tau^2
+\left(-1-\frac{2M}{r}+\frac{Q_c^2}{r^2}\right)^{-1}dr^2
+r^2(d\vartheta^2+\sinh^2\vartheta d\varphi^2).
\end{equation}
and its corresponding electromagnetic potential
\[
\Phi=e^{i\theta}\frac{Q_c}{r}
\]
after a trivial shift.
In this case the only constraint on the values of the parameters
$M$ and $Q_c$ is $M^2+Q_c^2\neq 0$. 
The range for the coordinate $r$ for
which the metric is static is given by
$-M-\sqrt{M^2+Q_c^2}<r<-M+\sqrt{M^2+Q_c^2}$.
This is the hyperbolic counterpart of the Reissner-Nordstr\"om
solution.

\subsection{Case $c^2 < 1$}

The equation that differs from the previous case  $c^2 > 1$
is 
\begin{equation}
\label{eq:2ricciV}
\RR_{ab}=-2V_{,a}V_{,b}.
\end{equation}
We proceed in an analogous way to solve the system (\ref{eq:2ricciV}),
(\ref{eq:laplace}) and (\ref{eq:yorkV}). 
The difference in sign in (\ref{eq:2ricciV}) compared to (\ref{eq:1ricciV})
only affects the equation for $W$, whose solution is given now by
\begin{equation}
W=(Ae^{ix}+\cxc A e^{-ix})^{-2},
\label{eq:W}
\end{equation}
where $A$ is a complex number.
The same previous procedure shows now that the surfaces
$(S_2,\Omega_{AB})$ are of positive constant
curvature 
$4A\cxc A$.
Coordinates $\{\vartheta,\varphi\}$ can therefore be chosen such that
\[
\Omega_{AB}dx^A dx^B=
\frac{1}{4A\cxc A}(d\vartheta^2+\sin^2\vartheta d\varphi^2),
\]
where $\vartheta\in(0,\pi)$ and $\varphi\in[0,2\pi)$. 
The complete line-element of the solution is found  using (\ref{eq:W})
on (\ref{eq:ds2V}) and taking into account (\ref{eq:k3}) for
$V=x$.
This case is analogous to the case (ii) above.
When $A+\cxc A\neq 0$, as expected,
the change of coordinates
\[
e^{2ix}=\frac{i-4\sqrt{A\cxc A(1-c^2)}\cxc A r}{i+4\sqrt{A\cxc A(1-c^2)}Ar},\quad
\tau=2\epsilon \frac{\sqrt{A\cxc A(1-c^2)}}{A+\cxc A}t,
\]
and the renaming
\[
Q\equiv\frac{\epsilon}{4 A\cxc A\sqrt{1-c^2}},\quad
M\equiv\frac{i(A-\cxc A)}{8(A\cxc A)^{3/2}\sqrt{1-c^2}}
\]
is what takes us to the Reissner-Nordstr\"om metric (\ref{eq:R-N}),
but for $M^2-Q^2<0$. Note that
with the above definitions $4A\cxc A(M^2-Q^2)=-(A+\cxc A)^2 Q^2$.

If $A+\cxc A= 0$ the change 
$z=a\cot x$ with $a^{-1}\equiv -(A-\cxc A)^2\sqrt{1-c^2}$
and $T=\sqrt{1-c^2}t$, which induces the change
$\Psi\to\Psi/\sqrt{1-c^2}$, leads to
\begin{equation}
ds^2=-\left(1+\frac{z^2}{a^2}\right)dT^2+
\left(1+\frac{z^2}{a^2}\right)^{-1} dz^2+a^2
(d\vartheta^2+\sin^2\vartheta d\varphi^2),
\label{eq:ds2_iv_1}
\end{equation}
and the electromagnetic potential (after a trivial shift) 
$
\Phi=-e^{i\theta}z/a.
$
The metric corresponds again to the near-horizon Bertotti-Robinson
spacetime (\ref{eq:ber-rob}), now in different coordinates $\{T,z\}$.

Let us stress the fact that the ``intrinsic'' difference that has led to
(\ref{eq:ber-rob}) and (\ref{eq:ds2_iv_1})
in the present setting
lies in the different
sign of the scalar curvature $\RR$ of the scaled quotient space $\hh_{ab}$
with respect to the Killing vectors $\partial_\tau$
and $\partial_T$, respectively, but it is not an intrinsic
property of the spacetime.
In other words, the difference lies in the possibility of choosing
timelike Killing vector fields in the Bertotti-Robinson spacetime with
associated positive and negative curved scaled quotient  spaces $\hh_{ab}$.
Note, however, that in the Reissner-Nordstr\"om case
the $\partial _t$ Killing is intrinsically defined (unit at infinity)
and that the sign of $\RR$ corresponds to the sign of $M^2-Q^2$,
which leads to two globally different spacetimes.


\section{Results}
The combination of the above theorems
and the classification of the functionally dependent
conformastat electrovacuum solutions in Section \ref{sec:classLzero}
leads to the following final result:
\begin{theorem} 
\label{res:families}
All conformastat electrovacuum spacetimes either belong
to the Majumdar-Papapetrou class or correspond to
either
\begin{enumerate}
\item[] the Bertotti-Robinson conformally flat solutions (\ref{eq:ber-rob}),
\item[] the non-extreme exterior Reissner-Nordstr\"om solution (\ref{eq:R-N}),
\item[] or its flat (\ref{eq:ds2i}) or hyperbolic (\ref{eq:pseuR-N}) counterparts.
\end{enumerate}\finn
\end{theorem}
Let us stress that the five classes are exclusive, and that the
extreme Reissner-Nordstr\"om case is included in the
Majumdar-Papapetrou class.
For completeness we include the Table \ref{tab:taula}
with a classification of the
conformastat electrovacuum solutions 
in terms of the geometrical properties of the timelike static congruence
defined by $\partial_t$ in (\ref{eq:stationary}) with $A_a=0$.
\begin{table}
\caption{\label{tab:taula}Possible quotient spaces  $(\Sigma_3,\hh_{ab})$
in conformastat electrovacuum solutions.}
\begin{indented}
\item[]\begin{tabular}{@{}llll}
    \br
    $\hh_{ab}$ Ricci scalar&$\Omega_{AB}$&&\\
    \mr
    $\RR=0$&-       &Majumdar-Papapetrou&\\
    $\RR>0$&flat      &Plane-symmetric fields&\\
           &spherical &Bertotti-Robinson&\\
           &          &Reissner-Nordstr\"om exterior& $M^2-Q^2>0$\\
           &hyperbolic&hyperbolic Reissner-Nordstr\"om&\\
    $\RR<0$&$\Rightarrow$spherical &Bertotti-Robinson&\\
           &          &Reissner-Nordstr\"om exterior& $M^2-Q^2<0$\\
\br
\end{tabular}
\end{indented}
\end{table}

The first corollary of this theorem and the classification
presented in Table \ref{tab:taula} constitutes an
improved local characterisation of Majumdar-Papapetrou.
The original local characterisation (see e.g. \cite{sols})
states that it is the class of
static electrovacuum spacetimes with \emph{flat} $\hh_{ab}$.
Here we have relaxed the requirement on $\hh_{ab}$ by showing that
\begin{corollary}
The Majumdar-Papapetrou class of solutions are the static
electrovacuum spacetimes with conformally flat $\hh_{ab}$
and $\RR=0$.
\end{corollary}

An alternative statement of the above theorem
is that \emph{the
static charged black hole related geometries, that is, the Majumdar-Papapetrou,
the Reissner-Nordstr\"om exterior and the near-horizon Bertotti-Roinson
geometry,
together with the trivial plane and hyperbolic generalisations of
Reissner-Nordstr\"om, are locally characterised by
being the only conformastat electrovacuum spacetimes}.
A global argument regarding asymptotic flatness can
then be used to establish that

\begin{corollary}
The conformastat electrovacuum asymptotically flat spacetimes are
either isometric
\begin{itemize}
\item to the asymptotically flat subset of the Majumdar-Papapetrou class
\item or to the exterior Reissner-Nordstr\"om solution.
\end{itemize}
\end{corollary} 

Further global considerations may be finally used to single out
the black hole geometries whithin the Majumdar-Papapetrou class,
the so-called \emph{standard} Majumdar-Papapetrou,
favoured by
the uniqueness results in \cite{chrus_tod_mBH}.
In order to do that one should ask for the global
requirements that single out the standard Majumdar-Papapetrou
among the complete class that appear as hypotheses in the results
shown in \cite{chrus_nadi}, which basically consist
of demanding a non-empty black hole region and a non-singular
domain of outer communications.

\ack
We are grateful to Marc Mars for his many suggestions, criticisms and
comments on this work. We also thank Jos\'e M M Senovilla for his
suggestions and careful reading of this manuscript.
All the computations shown in the appendices have been carried out
by programming the whole procedures in REDUCE.
GAG wants to thank the warm hospitality of Dept. de F\' isica Te\'orica,
UPV/EHU, where this work was made. RV is funded
by project IT-221-07 from the Basque Government,
and thanks support from project FIS2007-61800 (MICINN).

\appendix
\section{Proof of $\beta_{3}=0$}
\label{appendixA}
In this Appendix we present the
proof of Proposition \ref{res:beta3}, as indicated in
Section \ref{sec:Lnotzero}:
the proof that the integrability conditions of the equations
(\ref{eq:1}), (\ref{eq:2}) and (\ref{eq:ricci123}) imply an additional isometry.
This follows, exactly up to a couple of points and modulo some typos and
missing terms in intermediate steps, Sections 4 and 5 of \cite{perjes2}.
Let us recall that the two differencies of our proof with 
that in \cite{perjes2} come simply from the
two aspects in which the treatment
of the static electrovacuum case differs to that
of the stationary vacuum case, as explained in Section \ref{sec:commonframe}.

The first is the fact that our functions $\alpha$ and
$\gamma=\opp\alpha$ are two composed functions, one conjugate to the other,
and not one complex function
and its complex conjugate. The same goes for the coordinates
$x^1=\sigma$ and $x^2=\opp\sigma$. Although the product
$\alpha\gamma$ must be positive (see Subection \ref{sec:addsymm}),
other factors such as $\alpha_2 \gamma_1 (= \alpha_2 \opp{\alpha_2})$
can be negative in general. The positiveness of
$\alpha_2 \gamma_1 (= \alpha_2 \overline{\alpha_2})$ in the complex case
in used precisely in the final step of the proof in \cite{perjes2},
Section 5. Therefore we will need some further steps to complete the
proof in our case.

The second difference comes from the number $\nn$ (see (\ref{eq:ricciY})),
which infers a different numeric factor in one composed equation.
This difference will only imply different combinations to produce
the equations needed in each step of the proof.
We will indicate all the
calculations keeping an arbitrary $\nn$. The purpose is twofold. Apart
from the usual completeness reason, we also want
to reproduce the proof in \cite{perjes2}, and by doing so,
indicate (and fix) some intermediate errors (typos and some missing factors)
we have found in \cite{perjes2}, Section 5.
Therefore we will keep using the notation $\, \opp{}\,{} $
for the conjugate operation
that particularises to the complex conjugate in the complex case.

The starting point is the set of equations (\ref{eq:1}), (\ref{eq:2})
and the equations for the Ricci tensor (\ref{eq:ricci123}).
Note that $\nn$ only enters one equation in (\ref{eq:ricci123}),
the $(1,2)$ component.
The aim is to prove that $\beta$ does not depend on $x^3$.
To do so, we assume $\beta_3\neq 0$ in order to find a contradiction.
Recall that neither $\alpha$ nor $\gamma$ depend on $x^3$.
The first step is to strictly follow the arguments in \cite{perjes2}, Section 3,
where the integrability conditions for the functions $\beta$ and $\rho$
in the equations (\ref{eq:1}), (\ref{eq:2}) are obtained.
The integrability conditions eventually yield three differential
equations, namely (29b), (29c) and (29d) in \cite{perjes2}, together with
their conjugates, for the functions $\alpha$ and $\gamma$.

The second step follows Section 4 in \cite{perjes2}, in which
the equations for the Ricci tensor (\ref{eq:ricci123}) are used.
The generalisation to include an arbitrary $\nn$ is straightforward
and we simply indicate the equation involved and the result.
$\nn$ appears in the $\RR_{12}=\nn/(x^1+x^2)^2$ equation component,
and therefore contributes (only) to equation (31) --with (32)--
in \cite{perjes2}, which now reads
\begin{equation}
-(\RR_{11} +\rho^{-2} \frac{\gamma}{D}\RR_{33})D^{-1}
+ 4 \frac{\gamma}{D}(\frac{\alpha}{D}\RR_{11} + \frac{\beta}{D}\RR_{12})=
4 \frac{\beta\gamma}{D^2} \frac{\nn}{(x^1+x^2)^2}.
\label{eq:ricci12}
\end{equation}
This equation (and its conjugate) is convenient because,
after using the equations for the derivatives of $\rho$ and $\beta$
(equations (24) and (25) in \cite{perjes2}), provides the only combination
in which no $\partial_{x^3}$ derivatives appear, leading to a
polynomial of degree 9 in $\beta$. The 10 coefficients of the polynomial
must thus vanish, providing, in principle, 10 differential equations
for $\alpha$ and $\gamma$. Nevertheless, those 10 equations are proportional 
to two independent composed equations plus one imaginary equation.
Indeed, a straigforward calculation shows that the equations
corresponding to the odd powers of $\beta$ are all multiples of the composed equation
\begin{equation}
(\sigma+\opp{\sigma})^2\gamma_{12}-4(\sigma+\opp{\sigma})\gamma_{1}
-2(2\nn+1)\gamma=0.
\label{eq:new33}
\end{equation}
$\nn$ only affects the odd coefficients, and thus this is in fact the only equation
where $\nn$ appears. 
The equations for the even powers of $\beta$ provide the composed equation
\begin{equation}
\fl
(\sigma+\opp{\sigma})^2(\alpha\gamma_{11}+5\gamma\gamma_{22}-3\alpha_1\gamma_1-3\gamma_2^2)
+12\gamma^2+(\sigma+\opp{\sigma})(14\alpha\gamma_1-6\gamma_2)=0
\label{eq:34}
\end{equation}
plus the imaginary equation
\begin{equation}
(\sigma+\opp{\sigma})^2(\gamma_{22}-\alpha_{11})-6 (\sigma+\opp{\sigma})
(\gamma_2-\alpha_1)-12(\alpha-\gamma)=0.
\label{eq:29b}
\end{equation}
Equation (\ref{eq:new33}) particularises to (33) in \cite{perjes2}
for $\nn=1$, and (\ref{eq:34}) and (\ref{eq:29b}) correspond to
(34) and (29b) in \cite{perjes2} respectively. As claimed in \cite{perjes2},
the composed equation (\ref{eq:34}) implies (29d) in \cite{perjes2}
and one can easily check that (\ref{eq:new33}) implies (29c) in \cite{perjes2}.
As stated in \cite{perjes2}, there may appear another combination
of the equations for the Ricci tensor, namely
$\alpha\RR_{11}-\gamma\opp{\RR_{11}}(=\alpha\RR_{11}-\gamma\RR_{22})=0$.
However, this equation provides no new information.
All in all we are finally left with equations (\ref{eq:new33}),
(\ref{eq:34}) and (\ref{eq:29b}).

\subsection{The system of PDEs for $\alpha$ and $\gamma$}
Summing up, the complete system of equations for $\alpha$ and
$\gamma$ which decouples from the rest of the field equations is given by
(\ref{eq:new33}), (\ref{eq:34}) and (\ref{eq:29b}), which are conveniently rewritten
as
\begin{eqnarray}
  \label{eq:E1}
  E_1\equiv
  -\gamma_{12}+\frac{2}{r}\gamma_{1}+\frac{M}{2} \frac{\gamma}{r^2}=0\\
  \label{eq:E2}
  E_2\equiv
  -\gamma_{22}+\alpha_{11}+\frac{3}{r}(\gamma_2-\alpha_1)+\frac{3}{r^2}(\alpha-\gamma)=0\\
  \label{eq:tE3}
  \teq{E}_3\equiv
  \alpha\gamma_{11}+5\gamma\gamma_{22}-3(\alpha_1 \gamma_1 + \gamma_2^2)+
  \frac{1}{r}(7\alpha\gamma_1-3\gamma\gamma_2)+\frac{3}{r^2}\gamma^2=0,
\end{eqnarray}
where $r\equiv(\sigma+\opp\sigma)/2$ and $M\equiv2\nn+1$.
Since we will be only interested in $M=3$ and $M=9$
we will implicitly assume at some points
that certain polynomials in $M$
with other roots do not vanish, and in fact, that $M>0$. 
The $\teq{}$ accent is used here to keep an analogous notation
to that in \cite{perjes2}, and the only purpose is to denote differently
certain equations. Note, however, that the $\teq{}$ here
corresponds to the tilde in \cite{perjes2}.
Note that
$E_2=-\opp{E_2}$, and therefore the above system of equations contains
5 real equations.

The procedure consists of generating new
differential equations by computing the integrability
conditions of the system $(E_1,\opp{E_1},E_2,\teq{E}_3,\opp{\teq{E}_3})$.
This procedure will be fixed by the use of very specific sets of rules,
which must be applied in strict order.
Before setting the rules, let us produce two useful combinations
after using $E_1$ and $\opp{E_1}$ to eliminate $\gamma_{12}$
and $\alpha_{12}$ respectively:
\begin{eqnarray}
  \label{eq:E3}
\fl
  E_3\equiv 5\gamma E_2 + \teq{E}_3
  =  \alpha \gamma_{11}+5\gamma\alpha_{11}
  -3(\alpha_1 \gamma_1+\gamma_2^2)\nonumber\\
  +\frac{1}{r}(7\alpha\gamma_1
  +12\gamma\gamma_2 -15 \gamma\alpha_1)+ \frac{\gamma}{r^2}(15\alpha-12\gamma)=0,\\
  \label{eq:E4}
\fl
  E_4\equiv \partial_2 E_3+\left(\frac{2}{r}\gamma-6\gamma_2\right)E_2
  -\frac{2}{r}\teq{E}_3
  =\alpha_2 \gamma_{11}-\gamma_2 \alpha_{22}+\frac{1}{r}(2\gamma\alpha_{11}
  +3\alpha_1\gamma_2+\alpha_2 \gamma_1)\nonumber\\
  ~~~~+\frac{1}{r^2}\left(\frac{2M+3}{2}\gamma\alpha_1
    -\frac{2M+9}{2}\alpha\gamma_1 - 3\alpha\gamma_2\right)
    -(2M+9)\frac{\alpha\gamma}{r^3}=0.
\end{eqnarray}
The first and main rule is
\begin{enumerate}
\item [$(i)$] Multiplication by unknown functions (or their derivatives)
is allowed only when the resulting equation does not exceed the cubic
degree in the unkown functions.
\end{enumerate}
This rule only affects the choice
of combinations to generate new equations. Since we are going to
indicate these combinations explicitly, this rule does not need to be
implemented in the algorithm. It must also be stressed that in all the equations
the factors that will be isolated (and thence ``eliminated'') appear
linearly and with a non-zero multiplicative factor.     
The first set of rules, as such,
$\setrules_1=\{(ii),(iii),(iv),(v),(vi)\}$ reads:
\begin{enumerate}
\item [$(ii)$] eliminate $\gamma_{12}$ and $\alpha_{12}$ using
  $E_1$ and $\opp{E_1}$ respectively.
\item [$(iii)$] eliminate the product $\alpha_2\gamma_{11}$ by using $E_4$.
\item [$(iv)$] eliminate the product $\gamma \alpha_{22}$ by using $\opp{\teq{E}_3}$.
\item [$(v)$] eliminate the product $\alpha \gamma_{11}$ by using $\teq{E}_3$.
\item [$(vi)$] eliminate $\gamma_{22}$ by using $E_2$.
\item [$(vii)$] eliminate the product $\gamma_1\alpha_{22}$ by using
rule $(vi)$ applied to $\opp{E_4}$.
\end{enumerate}
In what follows we simply indicate the chain of equations used,
and the explicit expressions will be only given when needed.
For the sake of concreteness we prefer to specify whenever any set of rules
is applied to any expression $f$ by $\setrules(f)$.

The sequence of equations starts with
\[
E_5\equiv\setrules_1[~\gamma\partial_2 E_4~],
\]
and follows with
\begin{eqnarray*}
\fl
\teq{E}_6 \equiv \setrules_1[~ r^2\partial_2 E_5-2 r E_5-12\alpha_2 E_3 ~],\\
\fl
E_6 \equiv \setrules_1[~ \frac{3}{51-2M}\teq{E}_6+\frac{3}{2}
(\gamma \opp{E_4}-\alpha_1 \opp{E_3}) ~],\\
\fl
E_7 \equiv  r^2 \frac{2}{3(33-M)} \setrules_1[~
2\partial_2\teq{E}_6-\frac{4}{r}\teq{E}_6
+(M-18)\partial_2(\setrules_1[ \gamma_1\opp{E_3} ])
-\frac{30}{r}\gamma_1\opp{E_3}
~].
\end{eqnarray*}
Note that in the third
factor, as indicated, one must apply some
rules before differentiating.
The chain of equations follows with
\begin{eqnarray*}
\fl
E_8\equiv \setrules_1[~
\frac{1}{2}\partial_2 E_7 - \frac{7M + 27}{6}\alpha_2 E_3+\frac{(5M+18)(2 M-51)}{9(33-M)} \opp{E_6}
- \frac{1}{r}E_7
~],
\end{eqnarray*}
which results in a first order equation. From this point onwards
it is convenient to define a new set of rules (keeping the first rule (i)):
$\setrules_2=\{(ii),(vii),(iii),(iv),(vi)\}$.
The chain follows with
\[
  E_9\equiv\setrules_2[~ \partial_1 \teq{E}_8 ~],
\]
and a new sequence given by
\begin{eqnarray*}
  F_5 \equiv \setrules_2[~ \gamma_1\partial_2 E_4 ~],\\
  F_6 \equiv \setrules_2[~ \partial_2F_5-\frac{2}{r}F_5 ~],\\
  F_9 \equiv \setrules_2[~ \partial_2\opp{E_8} ~],\\
  F_7 \equiv \setrules_2[~ \partial_2 E_6-\frac{2}{r}E_6 ~],
\end{eqnarray*}
used to construct
\begin{eqnarray*}
\fl
E_{10} \equiv \setrules_2[~
4\partial_2 E_9+\partial_2(r^2 F_6)-2rF_6-\frac{1}{8}E_9-\frac{4}{r}F_9
~]\\
\fl
E_{11} \equiv \setrules_2[~
2 \partial_2(r^2E_{10})-4r E_{10}
+4(13M+56)\left(-F_9+\frac{2}{r}\opp{E_8}\right)\\
-\frac{(2M^2-M-141)(2M-51)}{6(M - 33)}\opp{F_7}
-\frac{M+2}{4}(M^2+10M+5)\frac{1}{r^2}\alpha E_3\\
+\frac{(9M^3 + 106M^2 + 645M + 876)}{4(M + 5)}\frac{1}{r^2}E_7
~].
\end{eqnarray*}
The general explicit
expression of $E_{11}$ reads
\begin{eqnarray*}
\fl
-24r^2(M+5) E_{11} =
\fa^{2} \gamma \left(322 M^{3}+3771 M^{2}+12018 M+7713\right)\\
+2 \fa \gamma_1 \alpha \left(319 M^{3}+4560 M^{2}+19473 M+24888\right)\\
+\fc^{2} \alpha 
\left(148 M^{3}+7797 M^{2}+64416 M+147303\right)\\
+2 \fc \alpha_2 \gamma \left(346 M^{3}+3831 M^{2}+14262 M+20433\right).
\end{eqnarray*}
The procedure follows by taking the imaginary part
\[
\teq{E}_{11}\equiv \frac{4}{3(33-M)(M+1)} r^2 (E_{11}-\opp{E_{11}})
\]
from where
\begin{eqnarray*}
\fl
E_{12} \equiv  \setrules_2[~ \partial_1 \teq{E}_{11} ~],\\
\fl
F_{12} \equiv   \setrules_2[~ \partial_2 \teq{E}_{11} ~],\\
\fl
\teq{E}_{13} \equiv   \setrules_2[~ 2r^2\left(
  \partial_2 E_{12}-\frac{2}{r}(E_{12}+F_{12})
  \right)
  -\frac{23M^2 - 236M - 483}{9(M + 1)}\teq{E}_{11} ~].
\end{eqnarray*}
$\teq{E}_{13}$ reads, explicitly,
\begin{eqnarray*}
E_b \equiv \frac{27(M+1)^2}{8(55M^3  + 1187M^2   + 6087M + 8091)}\teq{E}_{13}=\\
\gamma\fa^2-\alpha\fc^2=0.
\end{eqnarray*}
The next equation is given by
\[
\fl
E_a\equiv \frac{1}{3}(\teq{E}_{11} - \frac{29M+141}{3(M+1)} E_b)=
 \fc \alpha_2 \gamma - \fa \gamma_1 \alpha =0.
\]

Let us recall here that the
only positiveness property we can use in the general case
is $\alpha\gamma>0$.

Let us set up a new set of rules $\setrules_{ab}=\{(b),(a)\}$, where
\begin{enumerate}
\item [$(b)$] eliminate the factor $\alpha\gamma_2^2$ using $E_b$,
\item [$(a)$] eliminate the factor $\alpha_2\gamma_2\gamma$ using $E_a$,
\end{enumerate}
which applied to $E_{11}$ leads to
\begin{equation}
r^2(M+5)\setrules_{ab}[~ E_{11} ~]
=\fa\left[A\gamma\fa  + 2B\gamma_1 \alpha\right],
\label{eq:68}
\end{equation}
with
\begin{eqnarray}
&&B=\frac{1}{24}\left(-665 M^{3}-8391 M^{2}-33735 M-45321\right),\nonumber\\
&&A=\frac{1}{12}\left(-235 M^{3}-5784 M^{2}-38217 M-77508\right).
\label{eq:AB}
\end{eqnarray}
Note that $A<0$ and $B<0$ since $M>0$.
The factor $\fa$ cannot vanish, since otherwise
$\opp{E_1}$ would lead to $\alpha=0$,
and the same argument holds for the factor $\fc$ using $E_1$ and
$\gamma\neq 0$.
As a result, (\ref{eq:68}) and its conjugate
lead to the next pair of equations:
\[
E_0\equiv A\alpha\fc  + 2B\alpha_2 \gamma =0.
\]
We now use this equation to set up the next set of rules
$\setrules_{0}=\{(0i),(0ii)\}$ where
\begin{enumerate}
\item [$(0i)$] eliminate $\alpha \gamma_2$ using $E_0$,
\item [$(0ii)$] eliminate $\gamma \alpha_1$ using $\opp{E_0}$.
\end{enumerate}
The chain of equations follows with
\[
E_{14}\equiv \setrules_2[~ \partial_1 E_0 -\frac{2}{r}E_0 ~],~~~~
F_{14}\equiv \setrules_2[~ \partial_2 E_0 -\frac{2}{r}E_0 ~],
\]
from where we get
\[
\teq{E}_{14}\equiv \frac{1}{\alpha\gamma} \setrules_0[~ \alpha\gamma E_{14} ~],
\]
which explicitly reads
\begin{equation}
\label{eq:teq14}
\teq{E}_{14}=2 \alpha_2 \gamma_1 B \left(2 A^{
-1} B+1\right)+\alpha\gamma \frac{1}{r^{2}}\left(\frac{1}{2} A \nn+A+B \nn\right)=0.
\end{equation}
Note that $\teq{E}_{14}=\opp{\teq{E}_{14}}$.
Since $\alpha\gamma>0$ this equation implies $\alpha_2 \gamma_1<0$.
What we will really need later is simply $\alpha_2 \gamma_1\neq 0$.
From (\ref{eq:teq14}) we set up the new rule $\setrules_{14}=\{(14i)\}$, where
\begin{enumerate}
\item [$(14i)$] eliminate $\alpha_2\gamma_1$ using $\teq{E}_{14}$.
\end{enumerate}
The next equation reads
\[
F_{15}\equiv \setrules_0[~ F_{14} ~],
\]
which explicitly reads
\begin{eqnarray*}
F_{15}&=
&\alpha_{11}\alpha( A -10  B )+6 \alpha_1^{2} B +\alpha_2 \gamma_2 ( A + 8 B)\\
&&+\frac{1}{r}\left(3\alpha_1\alpha(2 B -A)
-2\alpha_2 \gamma(   A +8 B)\right)
+3\frac{1}{r^2} \alpha^{2} \left(A-2 B\right)=0,
\end{eqnarray*}
which, since $A -10  B>0$,
we use to set up the last rule $\setrules_{15}=\{(15i)\}$, where
\begin{enumerate}
\item [$(15i)$] eliminate $\alpha_{11}\alpha$ using $F_{15}$.
\end{enumerate}
The final step consists on using the previous $E_a$, differentiate it,
and use the sets of rules we have just defined in a very specific order.
The precise algorithm starts with
\[
E_{16}\equiv
\gamma \setrules_{15}[~ \alpha \setrules_{1}[~ \partial_2 E_a ~] ~].
\]
Note that at this point we have ignored rule (i), but the outcome will be
precisely the desired result, because
\[
E_{final}\equiv \setrules_{14}[~ \setrules_{0}[~
\setrules_{14}[~ \setrules_{ab}[~ E_{16} ~] ~] ~] ~]
\]
reads, explicitly,
\[
\fl
E_{final}=\gamma_1 \alpha^{3} \gamma 
\frac{1}{2r^{2}}\frac{4 \left(
\left(M-22\right) A-2 B M\right) B^{2}+\left(3 A M+6 A+10 B M+16 B\right) A^{2
}}{A\left(A+2 B\right)\left(10 B-A\right)}=0.
\]
The last factor, after using (\ref{eq:AB}) to
introduce the values of $A$ and $B$ in terms of $M$,
is a fraction containing polynomials in $M$ in which all
the coefficients are positive numbers. Therefore, the only solution
to $E_{final}=0$ would be  $\gamma_1\alpha^{3} \gamma=0$, which is
not allowed by virtue of (\ref{eq:teq14}) and $\alpha\gamma>0$.

We have thus shown that  $\alpha\gamma>0\Rightarrow\beta_{3}=0$
for any positive $\nn$, and in particular, in the stationary vacuum case
($\nn=1$), recovering the result in \cite{perjes2},
and in the static electrovacuum case ($\nn=4$).



\section{Proof of Proposition \ref{teo:noL}}
\label{appendixB}

The starting point is 
equations
(\ref{eq:1}), (\ref{eq:2}) and (\ref{eq:ricci123})
when 
all functions depend only on $\sigma$ and $\opp\sigma$.
Equation (\ref{eq:2}) is used to isolate the two derivatives of $\rho$,
\begin{eqnarray}
2D \frac{\rho_{1}}{\rho}=\beta\beta_{,1}-\gamma\beta_{,2}
+\frac{1}{r}(\gamma-\beta)\beta,\nonumber \\
2D \frac{\rho_{2}}{\rho}=\beta\beta_{,2}-\alpha\beta_{,1}
+\frac{1}{r}(\alpha-\beta)\beta, \label{eq:drho}
\end{eqnarray}
where $r\equiv(\sigma+\opp\sigma)/2$.
These two equations, which are of course
related by conjugation, will be used to eliminate $\rho$ in what follows.
The integrability condition will be dealt with later.

From equation (\ref{eq:1}) and taking into account that
$\alpha\gamma>0$, we can now isolate $\alpha_2$, and its conjugate
$\gamma_1$, to obtain
\begin{eqnarray}
  \alpha_2=\gamma^{-1}\left[-2\beta \alpha_1
    + \alpha\gamma_2+\alpha\beta_1+\beta\beta_2
    +\frac{1}{r}\left(3\alpha\beta-2\alpha\gamma-\beta^2\right)\right],\nonumber\\
  \gamma_1=\alpha^{-1}\left[-2\beta \gamma_2
    + \gamma\alpha_1+\gamma\beta_2+\beta\beta_1
    +\frac{1}{r}\left(3\gamma\beta-2\alpha\gamma-\beta^2\right)\right].
\label{eq:a2c1}
\end{eqnarray}
We use these two expressions to compute the second derivatives
$\alpha_{12},\alpha_{22},\gamma_{11},\gamma_{22}$  in terms of
$\alpha_1$, $\gamma_2$, $\gamma_{12}$, $\alpha_{11}$,
and the first and second derivatives of $\beta$.
We use their substitutions in what follows.

We concentrate now on the Ricci equations (\ref{eq:ricci123}).
From the equation $\RR_{33}=0$ we isolate $\beta_{12}$, which reads
\begin{eqnarray}
2 D \beta_{12}&=&\gamma(\alpha_1+\beta_2)\beta_2+\alpha(\gamma_2+\beta_1)\beta_1
-\beta(\alpha_1\beta_1+\gamma_2\beta_2+2\beta_1\beta_2)\nonumber\\
&&+\frac{\beta}{r}[(\beta-\gamma)\alpha_1+(\beta-\alpha)\gamma_2]
+\frac{\beta}{r^2}(\alpha\gamma+\beta^2-\beta\alpha-\beta\gamma).
\label{eq:subs_b12}
\end{eqnarray}
Now, from the real combination $\alpha \RR_{11}+\gamma \RR_{22}=0$ we isolate
$\beta_{11}$, which yields
\begin{eqnarray}
  &&\fl
  \alpha D \beta_{11}=\gamma(\alpha\gamma_2-\beta\alpha_1)\beta_2
  -\beta[\alpha(\beta_1+\gamma_2)\gamma_2
  +(\gamma\alpha_1-2\beta\gamma_2)\alpha_1]
  +(2\alpha\gamma-\beta^2)\alpha_1\beta_1\nonumber\\
  &&\fl\quad + \frac{1}{r}
  \left[(\gamma-\beta)\alpha(3\beta\gamma_2-2\gamma\beta_2)
    +\gamma\beta(2\alpha\beta_1-3\beta\alpha_1)
    +(\beta^2-3\alpha\gamma)\alpha\beta_1+(2\alpha\gamma+\beta^2)\beta\alpha_1
  \right]\nonumber\\
  &&\fl\quad +\frac{\alpha}{r^2}\left[
    \gamma(5\beta^2-\alpha\gamma-2\beta\gamma)
    -\frac{1}{2}\beta(\alpha\gamma+3\beta^2)
  \right].
\label{eq:b11}
\end{eqnarray}
The complex cojugate equation provides $\beta_{22}$, which solves
in turn the imaginary equation $\alpha \RR_{11}-\gamma \RR_{22}=0$.
This equation is in fact equivalent to the compatibility condition of the
above system (\ref{eq:drho}) for $\rho$.
It is straightforward to check that the compatibility condition
$\beta_{112}=\beta_{121}$ is automatically satisfied.
The first important consequence of (\ref{eq:b11}) is that
if $\beta=0$ the equation reduces to $\gamma=0$, which contradicts
$\alpha\gamma>0$. We must therefore take $\beta\neq 0$ in what follows.

The equation $\RR_{13}=0$ is identically satisfied, so it only remains
to consider the equation $\RR_{12}=\nn/(4r^2)$.
It is convenient first to substitute $\alpha_1$ and $\gamma_2$ by 
two new functions $Z$ and $W\equiv\opp Z$ defined by the relations
\begin{eqnarray}
  \alpha_1=\frac{1}{4\beta}
  \left[2\beta\beta_2+3\alpha\beta_1+\frac{5}{r}\alpha\beta
    -\frac{2}{r}\beta^2+W
  \right],\nonumber \\
  \gamma_2=\frac{1}{4\beta}
  \left[2\beta\beta_1+3\gamma\beta_2+\frac{5}{r}\gamma\beta
    -\frac{2}{r}\beta^2+Z
  \right],
\label{eq:a1c2}
\end{eqnarray}
from where we will also obtain $\alpha_{11}$ and $\gamma_{12}$
in terms of $Z_1$ and $W_1$.
Let us   also use the substitutions
\[
b_1\equiv\beta_1-\frac{\beta}{r},\quad b_2\equiv\beta_2-\frac{\beta}{r}.
\]
The equation $\RR_{12}=\nn/(4r^2)$ thus leads to the real relation
\begin{equation}
  \label{eq:F}
  \fl
  F\equiv 9\alpha\gamma\left(\alpha b_1^2 + \gamma b_2^2-2\beta b_1 b_2
    -\frac{8 \beta}{9 r^2}(\nn+2)D\right) - \alpha Z^2 -\gamma W^2 + 2\beta WZ=0.
\end{equation}

\subsection{Case A:} Let us assume $Z\neq 0$, so that
$\iota^2(\alpha Z^2+ \gamma W^2 - 2\beta ZW)>0$ (this is a positive definite
product due to $\iota^2 D=\iota^2(\alpha\gamma-\beta^2)>0$).
The procedure now consists on finding new equations on
$Z$, $W$, $Z_1$ $W_1$, $b_1$, $b_2$ and $\alpha,\beta,\gamma$
by first differentiating $F$ and follow by using the substitutions above.
In the following expressions we also use
(\ref{eq:F}) by isolating $b_2^2$.
From the first two derivatives $\partial_1 F$
and $\partial_2 F$ one can isolate $Z_1$ and $W_1$ and find the
explicit expressions:
\begin{eqnarray}
  \label{eq:W1}
  \fl
  &&24\alpha\gamma\beta D\left(W_1-\frac{2}{r}W\right)=\nonumber\\
  &&\quad 3\alpha\gamma\left\{
    \left[(9\alpha\gamma-12\beta^2)W-\alpha\beta Z\right] b_1+
    \left[(5\alpha\gamma-4\beta^2)Z+3\gamma\beta W)\right] b_2  
  \right\}\nonumber\\
  &&\quad +(9\alpha\gamma-6\beta^2)\gamma W^2-(14\alpha\gamma-12\beta^2)
  \beta W Z+(5\alpha\gamma-6\beta^2)\alpha Z^2\nonumber\\
  &&\quad + 4(16-\nn) D \frac{\alpha^2\gamma^2\beta}{r^2}
  \frac{2(\alpha Z -\beta W)Z+3\alpha\gamma(Z b_2-W b_1)}
  {\alpha Z^2+ \gamma W^2 - 2\beta ZW},
\end{eqnarray}
\begin{eqnarray}
  \label{eq:Z1}
  &&\fl 24\alpha\beta D\left(Z_1-\frac{1}{r}Z\right)=
  (4\alpha\gamma+6\beta^2)W Z -3\gamma\beta W^2-7\alpha\beta Z^2\nonumber\\
  &&\fl \quad +3\alpha [3\gamma \beta W+(9\alpha\gamma-16 \beta^2) Z] b_1+
  3[(11\alpha\gamma-4\beta^2)\beta Z-(5\alpha\gamma-2\beta^2)\gamma W] b_2
  \nonumber\\
  &&\fl \quad + 4(16-\nn)\alpha\gamma D \frac{\beta}{r^2}
  \frac{2\alpha\beta Z^2 -2\alpha\gamma W Z -3\alpha^2\gamma Z b_1
  +3\alpha\gamma(2\beta Z -\gamma W) b_2}{\alpha Z^2+ \gamma W^2 - 2\beta ZW}.
\end{eqnarray}
We continue by taking the $\partial_2$ derivatives of (\ref{eq:a1c2})
and use the above expressions to obtain $Z_2$ and $W_2$ in terms of
$Z$, $W$, $b_1$, $b_2$ and $\alpha,\beta,\gamma$.
One can therefore investigate the compatibility condition
$Z_{12}=Z_{21}$ (equivalent to $W_{12}=W_{21}$),
which provides one real equation:
\[
  (16-\nn) (\gamma W-\beta Z) D^2 K F_{II}=0,
\]
where
\[
K\equiv \alpha Z^2 - 2\beta Z W + \gamma W^2,
\]
and
\begin{eqnarray}
  \label{eq:fii}
  \fl F_{II}\equiv 6 \alpha\gamma\left[K+\frac{16-\nn}{r^2}\alpha\gamma\beta D\right]
  \oppRe[b_1 \alpha(\beta Z-\gamma W)]+\nonumber\\
  \fl \quad +K \left[
    K(2 \beta^2-\alpha\gamma)-\alpha\beta \gamma\frac{D}{r^2}
    \left(3(\nn+8)\alpha\gamma+2(\nn+2)\beta^2\right)
  \right]\nonumber\\
  \fl \quad - 4 \left((14-\nn)\nn+32\right)\frac{D^2}{r^2}\alpha^3\gamma^3\beta^2.
\end{eqnarray}
For the cases we are interested in we can assume $\nn\neq 16$, and since
$\gamma W-\beta Z\neq 0$ as otherwise $K=0$, we necessarily have $F_{II}=0$.
In the complex case (stationary vacuum) studied in
\cite{perjes3,perjes3p} one resorts to the fact that
$\beta\equiv (\sigma+\cxc\sigma)^2(\yy\cdot\cxc\yy)>0$
and $\nn=1$ to establish that
$K+\frac{1}{r^2}(16-\nn)\alpha\beta\gamma D>0 $.
In the general case, however, one must still consider two subcases.

\textit{Subcase A1:} $K+\frac{1}{r^2}(16-\nn)\alpha\beta\gamma D\neq 0$.
We use (\ref{eq:fii}) to isolate $b_1$ and consider
the imaginary combination
$\oppIm[(\beta Z-\gamma W)\alpha \partial_1]
F_{II}=0$, which leads to
\[
\alpha\beta\gamma(\gamma W-\beta Z)DK^2 G_I=0,
\]
where the factor $G_I$ satisfies
\[
\fl
\frac{1}{r^4}G_I-W F_{II}=-\alpha
(\gamma W-\beta Z)\left(K+\frac{16-\nn}{r^2}\alpha\beta\gamma D\right)
2\oppIm (3\alpha\gamma b_2 Z-\alpha Z^2).
\]
(Note that for this last step one must not use $F_{II}$ explicitly
and leave $b_1$ unsubstituted.) As a result, the equation
\begin{equation}
  \label{eq:fiii}
  F_{III}\equiv\oppIm (3\alpha\gamma b_2 Z-\alpha Z^2)=0
\end{equation}
follows.
We proceed with a twin combination to the previous,
$\oppIm[(\alpha Z-\beta W)\alpha \partial_1]F_{II}=0$, to finish
recovering the two derivatives of $F_{II}$. This combination,
after neglecting non-vanishing terms, leads to $G_{III}=0$, where
$G_{III}$ satisfies
\[
\frac{1}{r^4}G_{III}-Z F_{II}=6\alpha\gamma(\gamma W-\beta Z)
\left(K+\frac{16-\nn}{r^2}\alpha\beta\gamma D\right)\oppIm(\alpha Z b_1),
\]
and therefore yields to
\begin{equation}
\label{eq:twin}
\oppIm(\alpha Z b_1)=0.
\end{equation}
If $\gamma W^2- \alpha Z^2 \neq 0$ equations (\ref{eq:fiii}) and
(\ref{eq:twin}) lead to $b_1=W/(3\alpha)$ (and $b_2=Z/(3\gamma)$),
which substituted on $F=0$ (\ref{eq:F})
implies $D\alpha\beta\gamma(\nn+2)=0$, which is impossible in the present case.
Since we are interested in the cases $\nn=1$ and $\nn=4$ we will also
assume in the following that $N+2\neq 0$.
Therefore we need
\begin{equation}
\label{eq:317}
\gamma W^2- \alpha Z^2 = 0
\end{equation}
to make (\ref{eq:fiii}) and
(\ref{eq:twin}) linearly dependent.


We continue by taking the imaginary combination
$(Z \partial_2-W \partial_1)$(\ref{eq:317})
and substituting $b_2$ from (\ref{eq:twin}). Using (\ref{eq:317}),
and after neglecting non-vanishing terms,
that combination is shown to lead to the following equation
\begin{eqnarray}
  \label{eq:318}
  \left[(15\alpha\gamma-3\beta^2)\alpha Z^2
    + (9\beta^2-21 \alpha\gamma)\beta W Z
    +6(16-\nn)\alpha^2\gamma^2\beta\frac{D}{r^2}\right]\alpha b_1\nonumber\\
  +\left[(5\alpha\gamma -9\beta^2)(\alpha Z^2-\beta W Z)
    +4(16-\nn)\alpha^2\gamma^2\beta\frac{D}{r^2}\right]W=0.
\end{eqnarray}
The real combination $(Z \partial_2+W \partial_1)$(\ref{eq:317})
is proportional to $\gamma W^2- \alpha Z^2$ and therefore bears no
information.

On the other hand, let us
take $\partial_1$(\ref{eq:fiii}) and apply the following
chain of substitutions: first $Z_1$ and $W_1$ from (\ref{eq:Z1}) and
(\ref{eq:W1}), followed by $\alpha_2$ and $\gamma_1$ from (\ref{eq:a2c1}),
then use (\ref{eq:a1c2}) and follow by first
substituting the first derivatives of $\beta$ by $b_1$ and $b_2$
and then the first derivatives of $b_1$ and $b_2$
by the corresponding expressions in terms of $Z,W,b_1,b_2,\alpha,\gamma,\beta$
that come from the above relations for
$\beta_{12}$, $\beta_{11}$ (and $\beta_{22}$).
Next substitute $b_2$ from (\ref{eq:fiii}) and
use the combination of (\ref{eq:F}) with (\ref{eq:twin})
so that $b_1^2$ can be isolated in terms of  $Z,W,\alpha,\gamma,\beta$ only.
Finally, use (\ref{eq:317}) to eliminate first the factor $\gamma W^2$
and then to express the equation in the form $f b_1 + g W=0$
for some factors $f$ and $g$ not depending on $b_1$,
just like equation (\ref{eq:318}).
At this point it is convenient to introduce the definition
\[
K\equiv \alpha Z^2+\gamma W^2-2\beta Z W,
\]
so that $K=2Z(\alpha Z-\beta W)$ because of (\ref{eq:317})
and use it within the factors $f$ and $g$
to express first $Z^2 W^2$ in terms of $K^2$ and $ZWK$
and then $Z^2$ and $W^2$ (separately) in terms of $K$ and $W Z$. 
Using this procedure the expression for $\partial_1$(\ref{eq:fiii})
can be cast as
\begin{eqnarray}
  \fl  -3\left[(2\alpha\gamma+\beta^2)K^2 -2 \beta DZWK\right] b_1\alpha
  \nonumber\\
  \fl + \left\{
    (5\beta^2-2\alpha\gamma)K^2-2\beta DZWK\right.\nonumber\\
  \fl \quad +4\alpha\gamma\beta\frac{D}{r^2}
    \left[
      -K\left((10-4\nn)\alpha\gamma-(2+\nn)\beta^2\right)-4(\nn+2)\beta DZW
    \right]\nonumber\\
  \fl \quad \left.
    +\frac{8(\nn+2)(16-\nn)}{r^4}\alpha^3\gamma^3\beta^2 D^2
  \right\} W=0.
\label{eq:A6}
\end{eqnarray}

We already have the equations needed to end the proof: (\ref{eq:fii}),
(\ref{eq:318}) and (\ref{eq:A6}). On top of the above defined $K$,
we will now make use of the following extra useful definitions
\[
\aa\equiv \frac{16-\nn}{2+\nn}\neq 0\quad (\mbox{and}\neq -1),\quad
\trian\equiv (2+\nn)\alpha\gamma\beta\frac{D}{r^2},
\]
so that in the stationary vacuum case $\aa=5$ and in
the static electrovacuum case $\aa=2$.
Let us stress that in the general case $\beta$
does not have a fixed sign, and thence neither $\trian$ has,
even for $\nn>2$.
It is only the complex case that ensures us that $\beta>0$
and therefore $\trian<0$ for $\nn>2$
(recall that $\alpha\gamma>0$, $\iota^2 D>0$).

After using (\ref{eq:twin}) to get rid of $b_2$
and (\ref{eq:317}) together with the above procedure for
expressions of the form $f b_1 + g W=0$
so that $f$ and $g$ depend on $W$ and $Z$ only through the factors
$K$ and $ZW$, equation (\ref{eq:fii}) reads
\begin{equation}
  \label{eq:321}
  \fl
  3\alpha^2 \gamma K (K+\aa\trian) b_1
  +\left[
    \alpha\gamma(K+4\trian)(K+\aa\trian)-2\beta^2K(K-\trian)
  \right] W=0.
\end{equation}
Analogously, equations (\ref{eq:318}) and (\ref{eq:A6}) read,
respectively,
\begin{equation}
  \label{eq:322}
  \fl
  3\left[
    (5\alpha\gamma-\beta^2)K - 4 DZWK + 4 \aa\trian \alpha\gamma
  \right] \alpha b_1
  +\left[(5\alpha\gamma-9\beta^2)K+8\aa\trian\alpha\gamma\right]W=0,
\end{equation}
\begin{eqnarray}
  \label{eq:323}
  \fl
  -3\left[
    (2\alpha\gamma+\beta^2)K^2-2 \beta DZWK
  \right]\alpha  b_1-\left\{
    (2\alpha\gamma-5\beta^2)K^2+2\beta DZWK\right.\nonumber\\
  \left.
    - 4\trian\left[
      \left((\aa-3)\alpha\gamma-\beta^2\right)K-4\beta DZW
    \right]
    -8\aa\trian^2\alpha\gamma
   \right\}W=0.
\end{eqnarray}
Let us now rewrite (\ref{eq:F}) conveniently as
\begin{equation}
\label{eq:new_f}
9\alpha^2 b_1^2K=W^2(K+8\trian).
\end{equation}
Since $\iota^2K>0$ this equation implies $\iota^2(K+8\trian)\geq 0$.
For $K+8\trian= 0$ it is necessary and sufficient that $b_1=0(=b_2)$.
In that case, though, (\ref{eq:321}) together with (\ref{eq:322}) lead to
$\aa=4$ ($N=8/5$). Since we are not interested in that case we can assume
$\aa\neq4$ in the following, so that $b_1\neq 0$ and thus
$\iota^2(K+8\trian)>0$.

The combination of (\ref{eq:321}) and (\ref{eq:322}) that cancels
the terms $\trian b_1$ reads
\begin{equation}
  \label{eq:324}
  \fl
  -3\alpha b_1 K (DK-4\beta DZW)
  +\left\{
    -K^2D+4[2\beta^2-(\aa-4)\alpha\gamma]\trian K+16\aa \trian^2 \alpha\gamma
  \right\}W=0.
\end{equation}
The combination (\ref{eq:324})$-$(\ref{eq:323}) leads to
\begin{equation}
  \label{eq:326}
\fl
  9K^2 \alpha b_1(\alpha\gamma+\beta^2)+
  \left[
    3(\alpha\gamma-3\beta^2)K^2+4\beta DZW(K+8\trian)
    +4\alpha\gamma K\trian(\aa-2)
  \right] W=0.
\end{equation}
On the other hand, let us isolate $WZ$ from (\ref{eq:324}), use
that on (\ref{eq:326}), multiply the result by $3K\alpha b_1$
and then use (\ref{eq:new_f}) to get rid of $b_1^2$.
Again, multiply the result by $b_1/(2W)$ and use
(\ref{eq:new_f}) to get rid of $b_1^2$. The resuling equation,
after neglecting the multiplying factors $W$ and $K+8\trian$, reads
\begin{equation}
  \label{eq:327}
  \fl
  -3\alpha b_1(K-4\trian)
  \left[
    (2\alpha\gamma+\beta^2)K+2\aa\trian\alpha\gamma
  \right]-K
  \left[
    2(\aa\alpha\gamma-2\beta^2)\trian+(2\alpha\gamma-5\beta^2)K
  \right]W=0.
\end{equation}
Another useful combination consists on taking (\ref{eq:321}),
multiply it by $\alpha b_1/W$ and use (\ref{eq:new_f})
to get rid of $b_1^2$ to get
\begin{equation}
  \label{eq:328}
\fl
  3\alpha b_1
  \left[
    (\alpha\gamma-2\beta^2)K^2 +(\alpha\gamma(\aa+4)+2\beta^2)K\trian
    +4\aa\alpha\gamma\trian^2
  \right]
  +\alpha\gamma(\aa\trian+K)(K+8\trian)W=0.
\end{equation}

Now, the combination
$-4\times$(\ref{eq:321})$-$(\ref{eq:327})$+2\times$(\ref{eq:328})
multiplied by $1/(3K\beta^2)$ leads to
\begin{equation}
  \label{eq:329}
  3\alpha b_1 K-(K-4\trian)W=0.
\end{equation}
Proceeding once more by multiplying this equation by $b_1$ and using
(\ref{eq:new_f}) to eliminate $b_1^2$, we obtain
a different relation between $b_1$ and $W$:
\begin{equation}
  \label{eq:329_bis}
  3\alpha b_1(K-4\trian)- (K+8\trian)W=0.
\end{equation}
Finally, isolating $b_1$ from
the latter and substituting in (\ref{eq:329}) we finally obtain
\begin{equation}
  \label{eq:fi1}
  \trian W^2 (\trian-K)=0,
\end{equation}
which now implies $\trian=K$ because we are assuming $Z\neq 0$ and
$\beta\neq 0$. We only need now to isolate $b_1$ from (\ref{eq:329})
and substitute that onto (\ref{eq:321}) using also $\trian=K$
to obtain
\[
K W \alpha\gamma (\aa + 1)=0,
\]
which contradicts our assumption $Z\neq 0$ in the present case.
This finishes \textit{subcase A1}.

\textit{Subcase A2:} $K+\frac{1}{r^2}(16-\nn)\alpha\beta\gamma D= 0$.
With the above definitions this is $K + \aa\trian = 0 $.
We only have to go back to equation (\ref{eq:fii}) and express
it in terms of $K$ to obtain
\[
K\beta (\aa + 1)=0,
\]
which contradicts our assumption $Z\neq 0$ in the present case.
This finishes \textit{subcase A2}
and therefore \textit{case A} completely.

\subsection{Case B:} We deal now with $Z=0$.
Let us take $\alpha_1$
from (\ref{eq:a1c2}) and use it
on the first equation in (\ref{eq:a2c1}) to get
\begin{equation}
  \label{eq:28III}
  \gamma \alpha_2 - \alpha \gamma_2+\frac{1}{2}\alpha \beta_1
  +\frac{\alpha}{r}\left(2\gamma-\frac{1}{2}\beta\right)=0,
\end{equation}
followed by $\gamma_2$ from (\ref{eq:a1c2}) to obtain
\begin{equation}
  \label{eq:29iii}
  4\alpha_2 \beta r - 3\beta_2 \alpha r + 3\alpha\beta=0,
\end{equation}
or equivalently, $(\alpha^4\beta^{-3}r^6)_{,2}=0$.
The solution is thus of the form $\alpha^4\beta^{-3}r^6=\zeta^4(x^1)$
for some analytic (or hyperbolic analytic \cite{lambertpiette}) function
$\zeta(x^1)$. From this equation we have
\[
\alpha=\zeta \beta^{3/4} r^{-3/2},\quad
\gamma=\opp\zeta \beta^{3/4} r^{-3/2},
\]
which used back into (\ref{eq:28III}) leads first to
\[
\beta^{-3/4}r^{9/2}\left(\frac{\beta}{r^2}\right)_{,1}=-4\opp\zeta.
\]
The first thing this equation implies is that $\opp\zeta$ is real,
and thence, by the (generalised) Cauchy-Riemann equations
$\partial_1\oppRe(\zeta)=\partial_2\oppIm(\zeta)$,
$\partial_2\oppRe(\zeta)=\iota^2 \partial_2\oppIm(\zeta)$,
$\zeta$ must be constant. The result
\[
\beta=r^2(a r^{-2}+b)^4
\]
for real constants $a$ and $b$ thus follows.
Introducing this solution together with the above expression
for $\alpha$ (and $\gamma$) into (\ref{eq:subs_b12})
leads to $ab(a+b r^2)=0$, which contradicts $\beta\neq 0$.
This finishes \textit{case B} and therefore the proof.

\end{document}